\documentclass[twocolumn,aps,pra,longbibliography,superscriptaddress,nofootinbib,10pt]{revtex4-1}
\usepackage[latin1]{inputenc}
\usepackage{graphicx,dcolumn,bm}
\usepackage{subfigure}
\usepackage{multirow}
\usepackage{array}
\usepackage{arydshln}
\usepackage{color}
\usepackage[colorlinks,linkcolor=blue,citecolor=blue,hyperindex]{hyperref}
\usepackage{amsfonts}
\usepackage{amssymb}
\usepackage{amsmath}
\usepackage{latexsym}
\usepackage{simplewick}
\usepackage{epstopdf}
\usepackage{multirow}
\usepackage{float}
\usepackage[table]{xcolor}
\usepackage{multibib}
\usepackage{mathrsfs}
\usepackage[sans]{dsfont}
\usepackage[mathscr]{eucal}
\usepackage{epsfig}
\usepackage{soul}
\definecolor{Blue}{rgb}{0.0,0.0,1}
\definecolor{Red}{rgb}{1,0.0,0.0}
\definecolor{Green}{rgb}{0,0.5,0.0}
\usepackage{tikz}
\usetikzlibrary{decorations.pathmorphing}
\usetikzlibrary{arrows}
\usetikzlibrary{intersections,shapes.arrows}
\usetikzlibrary{calc}
\usetikzlibrary{quotes,angles}
\usepackage{nicefrac}
\usepackage{pgfplots}
\usepgfplotslibrary{fillbetween}
\pgfplotsset{compat=1.13,colormap={violetnew}{rgb=(0.293416, 0.0574044, 0.529412) rgb=(0.394818,0.233715,0.671945) rgb =(0.49622,0.410025,0.814477) rgb=(0.588672,0.567494,0.910066) rgb=(0.663226,0.687282,0.911765) rgb=(0.73778,0.807069,0.913465) rgb=(0.807267,0.861883,0.894034) rgb=(0.874222,0.884211,0.864039) rgb=(0.941176, 0.906538, 0.834043)}}
\usepgfplotslibrary{groupplots} 
\usepgfplotslibrary[groupplots] 
\usetikzlibrary{pgfplots.groupplots} 
\usetikzlibrary[pgfplots.groupplots] 
\usepgfplotslibrary{statistics}
\usepackage{pgfplotstable}
\tikzset{jumpdot/.style={mark=*,solid},excl/.append style={jumpdot,fill=white},incl/.append style={jumpdot,fill=black}}
\pgfdeclarelayer{background}
\pgfsetlayers{background,main}
\begin{document}

\title{Unified entropies and quantum speed limits for nonunitary dynamics}
\author{Diego Paiva Pires}
\affiliation{Departamento de F\'{i}sica, Universidade Federal do Maranh\~{a}o, Campus Universit\'{a}rio do Bacanga, 65080-805, S\~{a}o Lu\'{i}s, Maranh\~{a}o, Brazil}

\begin{abstract}
We discuss a class of quantum speed limits (QSLs) based on unified quantum ($\alpha,\mu$)-entropy for nonunitary physical processes. The bounds depend on both the Schatten speed and the smallest eigenvalue of the evolved state, and the two-parametric unified entropy. We specialize these results to quantum channels and non-Hermitian evolutions. In the first case, the QSL depends on the Kraus operators related to the quantum channel, while in the second case the speed limit is recast in terms of the non-Hermitian Hamiltonian. To illustrate these findings, we consider a single-qubit state evolving under the (i) amplitude damping channel, and (ii) the nonunitary dynamics generated by a parity-time-reversal symmetric non-Hermitian Hamiltonian. The QSL is nonzero at earlier times, while it becomes loose as the smallest eigenvalue of the evolved state approaches zero. Furthermore, we investigate the interplay between unified entropies and speed limits for the reduced dynamics of quantum many-body systems. The unified entropy is upper bounded by the quantum fluctuations of the Hamiltonian of the system, while the QSL is nonzero when entanglement is created by the nonunitary evolution. Finally, we apply these results to the XXZ model and verify that the QSL asymptotically decreases as the system size increases. Our results find applications to nonequilibrium thermodynamics, quantum metrology, and equilibration of many-body systems.
\end{abstract}

\maketitle


\section{Introduction}
\label{sec:00000000001}

The processing of quantum information involves the ability to successfully distinguish quantum states~\cite{QInfProcess_2016_MTomm,RevModPhys.90.035005}. In recent years, information science has established a modern language to address this issue, thus contri\-bu\-ting to the understanding of quantum metro\-lo\-gy~\cite{PhysRevLett.96_010401,T_th_2014,PhysRevA.94.010102}, quantum computing~\cite{PhysRevLett.74.4091,arXiv:1203.5813}, quantum thermodynamics~\cite{RevModPhys.83.771,Goold2016}, and quantum communication~\cite{NatPhoton_1_165_2007,arXiv:2011.04672}. The idea is to introduce information-theoretic quantifiers that capture the sensitivity of these states in response to some phase enco\-ding process or a certain quantum evolution protocol~\cite{Nielsen_Chuang_infor_geom,IntJQuantInf_125_7_2009}. Importantly, some of these quantifiers have a clear geome\-tric meaning in terms of distances, thus endowing the space of quantum states with Riemannian metrics that are contractive under dy\-na\-mi\-cal maps~\cite{PhysRevLett.72.3439,JMathPhys_55_075211,10.1116_1.5119961,Ingemar_Bengtsson_Zyczkowski}.

As a matter of fact, von Neumann entropy is one of the most remarkable information-theoretic quantifiers, and finds applications ranging from quantum information science~\cite{PhysRevA.54.3824,JStatPhys_133_1161,RevModPhys.81.1665,PhysRevResearch.2.013161,JStatMech_033102_2016} to condensed matter physics~\cite{PhysRevLett.101.010504,PhysRevB.102.094303,PhysRevB.105.125413}. Its classical analog, i.e., Shannon entropy~\cite{Shannon1948}, also plays a role in applied sciences~\cite{Zhou2013}. In turn, such quantities constitute particular cases of a broader set of information measures given by the R\'{e}nyi~\cite{Renyi1961} and Tsallis entropies~\cite{Tsallis1988}, and have found applications as resource measures of coherence and entanglement~\cite{PhysRevA.94.052336,0253-6102-67-6-631,1751-8121-50-47-475303,NJP_20_053058_2018}. So far, several generalizations for these two entropies have been proposed, e.g., Petz-R\'{e}nyi relative entropy~\cite{PETZ198657,DBLPjournals_qic_Audenaert01,10.1063_1.5007167}, sandwiched R\'{e}nyi relative entropy~\cite{10.1063.1.4838856,Datta_2014}, and $\alpha$-$z$-relative R\'{e}nyi entropy~\cite{10.1063.1.4906367}, some of them not fully compatible with data processing inequality~\cite{doi:10.1063/1.4838855}. In particular, the Petz-R\'{e}nyi relative entropy has been applied in the study of phase encoding protocols~\cite{PhysRevA.102.012429}, and quantum speed limits~\cite{PhysRevE.103.032105} in closed quantum systems, but the case of nonunitary dynamics still remains as an issue.

The quantum unified entropy (UQE), also known as unified-$(\alpha,\mu)$ entropy, stands out as another important information-theoretic quantifier~\cite{doi:10.1063.1.2165794_Hu_Ye}. In turn, UQE denotes a two-parametric, continuous, and positive information measure, particularly recovering both the R\'{e}nyi and Tsallis entropies as limiting cases~\cite{Rastegin2011}. It is worth mentioning that UQE encompasses a whole family of quantum entropies that can be readily recovered by setting the parameters $(\alpha,\mu)$. It is noteworthy that UQEs finds applications ranging from witnessing monogamy of entanglement in multipartite systems~\cite{Sanders_JPhysMath,SciRep_7_1122_2017,Kim_SciRep_8_2018,YYetal_LaserPhysLett_18_2021} to characterizing quantum channels~\cite{Rastegin_JPhysA_2012,JPhysAMathTheor_46_285301}, and in the study of complexity beyond scrambling~\cite{JHighEnergPhys_41_2018}. Furthermore, UQE stands as a particular case of the unified $(\alpha,\mu)$-relative entropy~\cite{IntJTheorPhys_50_1282}. 

Here we will address the interplay of unified quantum entropies and quantum speed limits for nonunitary physical processes. The idea is investigate the QSL under the viewpoint of this versatile information-theoretic quantifier. We consider the rate of change of UQE, and derive a class of quantum speed limits for general nonunitary evolutions. The bounds depend on the smallest eigenvalue of the quantum state, also being a function of the Schatten speed. We specialize these results to the case of quantum channels, and for non-Hermitian systems. In particular, we set the single-qubit state and present numerical simulations to support these theoretical predictions. Moreover, for closed many-body quantum systems, we find the QSL for the marginal states of the system is a function of UQE, and the variance of the Hamiltonian. Importantly, our main contribution lies on the derivation of QSLs and bounds on UQE for nonunitary dynamics.

The paper is organized as follows. In Sec.~\ref{sec:00000000002}, we review useful properties regarding UQEs. In Secs.~\ref{sec:00000000004} and~\ref{sec:00000000005}, we address the connection between UQE and the quantum speed limit for nonunitary dynamics. In Secs.~\ref{sec:00000000008} and~\ref{sec:00000000009}, we specialize these results to the case of quantum channels and dissipative systems described by non-Hermitian Hamiltonians, respectively. In Sec.~\ref{sec:00000000011}, we focus on the QSL for the reduced dynamics of multiparticle systems, and illustrate our findings by means of the XXZ model. Finally, in Sec.~\ref{sec:00000000013} we summarize our conclusions.


\section{Unified quantum entropies}
\label{sec:00000000002}

In this section, we briefly review the main properties of unified quantum entropies. Let us consider a quantum system with finite-dimensional Hilbert space $\mathcal{H}$, with $d = \dim\mathcal{H}$. The space of quantum states $\mathcal{S} \subset \mathcal{H}$ is a set of Hermitian, positive semidefinite, trace-one, $d\times d$ matrices, i.e., $\mathcal{S} = \{\rho \in \mathcal{H} \mid {\rho^{\dagger}} = \rho,~\rho\geq 0,~\text{Tr}(\rho) = 1\}$. The quantum unified $(\alpha,\mu)$-entropy (UQE) is defined as~\cite{doi:10.1063.1.2165794_Hu_Ye,Rastegin2011}
\begin{equation}
\label{eq:0000000000001}
{{\text{E}}_{\alpha,\mu}}(\rho) := \frac{1}{(1 - \alpha)\mu} \left\{ { {{\left[{{f}_{\alpha}}(\rho)\right]}^{\mu}} - 1 }\right\} ~,
\end{equation}
with
\begin{equation}
\label{eq:0000000000002}
{{f}_{\alpha}}(\rho) =   \text{Tr}\left({{\rho}^{\alpha}}\right) ~,
\end{equation}
where $\alpha\in(0,1)\cup(1,+\infty)$ and $\mu \in (-\infty,0)\cup(0,+\infty)$. Note that ${{f}_{\alpha}}(\rho)$ in Eq.~\eqref{eq:0000000000002} plays a role of $\alpha$-purity and denotes a real-valued, nonnegative function for all $\alpha$. Indeed, taking the spectral decomposition $\rho = {\sum_j}\, {p_j}|j\rangle \langle{j}|$ in terms of the basis of states ${\{|j\rangle\}_{j = 1,\ldots, d}}$, with $0 \leq {p_j} \leq 1$ and ${\sum_j}\, {p_j} = 1$, we find that ${{f}_{\alpha}}(\rho) = {\sum_j}\, {p_j^{\alpha}} \geq 0$.

The unified entropy is nonnegative, ${{\text{E}}_{\alpha,\mu}}(\rho) \geq 0$, and remains invariant under unitary transformations on the input state, i.e., ${{\text{E}}_{\alpha,\mu}}({V}\rho{V^{\dagger}}) = {{\text{E}}_{\alpha,\mu}}(\rho)$, with $V{V^{\dagger}} = {V^{\dagger}}V = \mathbb{I}$, for all $\alpha\in(0,1)\cup(1,+\infty)$ and $\mu \neq 0$. In addition, UQE satisfies the upper bound ${{\text{E}}_{\alpha,\mu}}(\rho) \leq {[(1 - \alpha)\mu]^{-1}} \, \{ { [\text{rank}(\rho)]^{(1 - \alpha)\mu}} - 1\}$, for all $\alpha \neq 1$ and $\mu \neq 0$~\cite{doi:10.1063.1.2165794_Hu_Ye}. Importantly, the latter bound reproduces the ine\-qua\-li\-ty $ S(\rho) \leq \ln \, [ \text{rank}(\rho)]$ for the von Neumann entropy in the limit $\alpha \rightarrow 1$ and $\mu \neq 0$, with $S(\rho) = -\text{Tr}(\rho\ln\rho)$~\cite{Hayden2002}. 

Furthermore, it has been verified that UQE satis\-fies the properties: (i) concavity, ${\sum_l}\,{p_l}{{\text{E}}_{\alpha,\mu}}({\rho_l}) \leq {{\text{E}}_{\alpha,\mu}}({\rho})$ for $0 < \alpha < 1$ and $0 \leq \mu \leq 1$, where $\rho = {\sum_l}\, {p_l}{\rho_l}$, with $0 \leq {p_l} \leq 1$ and ${\sum_l}\, {p_l} = 1$; (ii) subadditivity, ${{\text{E}}_{\alpha,\mu}}({\rho_1}\otimes{\rho_2}) \leq {{\text{E}}_{\alpha,\mu}}({\rho_1}) + {{\text{E}}_{\alpha,\mu}}({\rho_2})$ for $0 < \alpha < 1$ and $\mu < 0$ ($\alpha \geq 1$ and $\mu \geq 0$), where the ine\-qua\-lity is reversed for $\alpha > 1$ and $\mu < 0$ ($0 < \alpha < 1$ and $\mu > 0$); (iii) Lipschitz continuity $|{{\text{E}}_{\alpha,\mu}}({\rho_1}) - {{\text{E}}_{\alpha,\mu}}({\rho_2})| \leq {(\alpha(\alpha - 1))^{-1}} {\|{\rho_1} - {\rho_2}\|_1}$, where $\|\bullet \|_1$ is the trace distance; (iv) nondecreasing under projective measurements, ${{\text{E}}_{\alpha,\mu}}({\rho}) \leq {{\text{E}}_{\alpha,\mu}}(\Phi(\rho))$, with $\Phi(\rho) = {\sum_l}\,{\mathcal{M}_l}\rho{\mathcal{M}_l}$ for a given set $\{{\mathcal{M}_l}\}$ of measurement operators; and (v) triangle inequality $|{{\text{E}}_{\alpha,\mu}}({\rho_1}) - {{\text{E}}_{\alpha,\mu}}({\rho_2})| \leq {{\text{E}}_{\alpha,\mu}}({\rho_{12}})$ for $\alpha > 1$ and $\mu \geq 1/\alpha$, with the marginal states ${\rho_{1,2}} = {\text{Tr}_{2,1}}(\rho_{12})$~\cite{doi:10.1063.1.2165794_Hu_Ye,Rastegin2011,JPhysAMathTheor_46_285301}.

Next, we comment on some particular cases of the unified entropy. On the one hand, the Tsallis entropy ${{\text{H}}_{\alpha}}(\rho) = {{\text{E}}_{\alpha,1}}(\rho) = {{({1 - \alpha })}^{-1}}\left[{{f}_{\alpha}}(\rho) - 1\right]$ is reco\-ve\-red for $\mu = 1$, with $\alpha > 0$ and $\alpha \neq 1$. In par\-ti\-cu\-lar, for $\alpha = 2$ it reduces to the linear entropy ${{\text{H}}_{2}}(\rho) = 1 - \text{Tr}({{\rho}^2})$, a quantity measuring the mixedness of a given quantum state~\cite{PhysRevA.70.052309}. On the other hand, the R\'{e}nyi entropy ${{\text{R}}_{\alpha}}(\rho) = {{\text{E}}_{\alpha,0}}(\rho) = {{({1 - \alpha })}^{-1}} \ln\left[ {{f}_{\alpha}}(\rho)\right]$ is recovered for $\mu = 0$, with $\alpha > 0$ and $\alpha \neq 1$. In particular, $\alpha = 0$ implies the max-entropy ${{\text{R}}_{0}}(\rho) = \ln \, [{\text{rank}(\rho) }]$, while the min-entropy ${{\text{R}}_{\infty}}(\rho) = - \ln {{\|\rho\|}_{\infty}}$ is obtained in the limit $\alpha \rightarrow \infty$, where ${{\|\bullet\|}_{\infty}}$ is the infinity norm~\cite{Bathia_Rajendra}. Moreover, the case $\alpha = 2$ sets the second-order R\'{e}nyi entropy, also known as collision entropy~\cite{PhysRevA.85.012108,10.1063.1.4838856}, which finds application in the characterization of quantum information for Gaussian states~\cite{PhysRevLett.109.190502}. Finally, for $\alpha = 1$, UQE recovers the von Neumann entropy.

\section{Bounding UQE}
\label{sec:00000000004}

\begin{figure}[!t]
\begin{center}
\includegraphics[scale=1.]{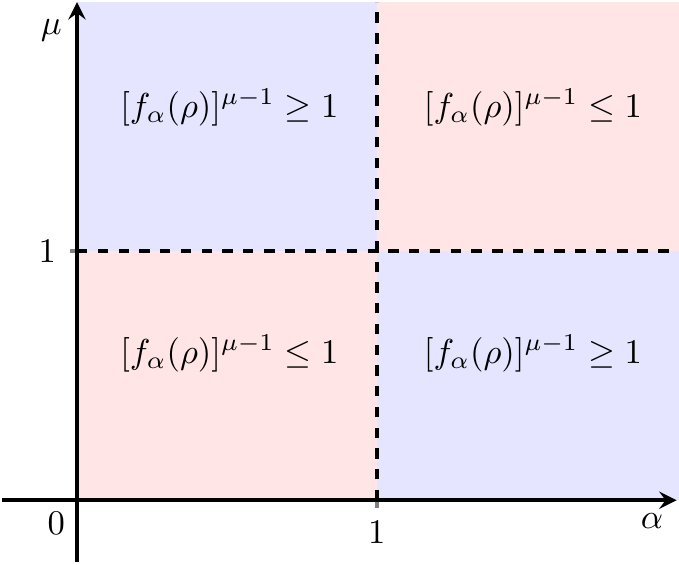}
\caption{(Color online) Phase diagram for the function $[{f_{\alpha}}(\rho)]^{\mu - 1}$, with $\rho\in\mathcal{S}$. On the one hand, the $\alpha$-purity sa\-tis\-fies the inequality ${f_{\alpha}}(\rho) \geq 1$ for $\alpha \in [0,1]$, and one finds $[{f_{\alpha}}(\rho)]^{\mu - 1} \leq 1$ for $0 < \mu < 1$, while $[{f_{\alpha}}(\rho)]^{\mu - 1} \geq 1$ for $\mu > 1$. On the other hand, the $\alpha$-purity behaves as ${f_{\alpha}}(\rho) \leq 1$ for $\alpha \geq 1$, which implies that ${[{f_{\alpha}}(\rho)]^{\mu - 1}} \geq 1$ for $0 < \mu < 1$, while we have that ${[{f_{\alpha}}(\rho)]^{\mu - 1}} \leq 1$ for $\mu > 1$.}
\label{fig:FIG01}
\end{center}
\end{figure}
The aim of this section is to derive an upper bound on UQE for a quantum state evolving under nonunitary dynamics. We consider a quantum system initialized at the state $\rho_0 \in \mathcal{S}$ undergoing a time-dependent nonunitary evolution ${{\mathcal{E}}_t}(\bullet)$, with $t \in [0,\tau]$. The evolved quantum state $\rho_t = {{\mathcal{E}}_t}({\rho_0})$ stands as a nonsingular, full-rank density matrix. Unless otherwise stated, from now on we will focus on the range $0 < \alpha < 1$ and $0 < \mu < 1$. For simplicity, we set $\hbar = 1$. The absolute value of the time derivative of unified quantum entropy ${{\text{E}}_{\alpha,\mu}}({\rho_t})$ read as
\begin{equation}
\label{eq:0000000000003}
\left|\frac{d}{dt}{{\text{E}}_{\alpha,\mu}}({\rho_t})\right| = \frac{1}{|1 - \alpha|} \, {{\left[{{f}_{\alpha}}({{\rho}_t})\right]}^{\mu - 1}} \left|\frac{d}{dt}{{f}_{\alpha}}({\rho_t})\right|  ~.
\end{equation}
Figure~\ref{fig:FIG01} shows a depiction of the phase diagram for the function ${[{f_{\alpha}}({\rho})]^{\mu - 1}}$ in the $\alpha$--$\mu$ plane. We point out that, for $\alpha \in (0,1)$, the $\alpha$-purity fulfills the lower bound ${f_{\alpha}}({\rho}) \geq 1$, with ${\rho} \in \mathcal{S}$. In this case, it follows that the condition ${[{f_{\alpha}}({\rho})]^{\mu - 1}} \leq 1$ is satisfied for $\mu \in (0,1)$~\cite{Rastegin2011}. Hence, applying such bound into Eq.~\eqref{eq:0000000000003} yields
\begin{equation}
\label{eq:0000000000004}
\left| \frac{d}{dt}{{\text{E}}_{\alpha,\mu}}({\rho_t}) \right| \leq \frac{\alpha}{\left| 1 - \alpha \right| } \left|\frac{d}{dt}{{f}_{\alpha}}({\rho_t})\right| ~.
\end{equation}
In Appendix~\ref{sec:00000000014} we prove that, for $\rho_t$ being a nonsingular density matrix, and $\alpha \in (0,1)$, the absolute value of the time derivative of $\alpha$-purity satisfies the inequality
\begin{equation}
\label{eq:0000000000005}
\left| \frac{d}{dt}{f_{\alpha}}({\rho_t}) \right| \leq \left({\kappa_{\text{min}}}({\rho_t}) + 1 - \alpha \right) {({\kappa_{\text{min}}}({\rho_t}))^{\alpha - 2}} \, {\left\|\frac{d{\rho_t}}{dt}\right\|_1} ~, 
\end{equation}
where ${\| {\hat{\mathcal{O}}}  \|_p} := (\text{Tr}\,[( {\mathcal{O}^{\dagger}}\mathcal{O})^{p/2}])^{1/p}$ denotes the Schatten $p$-norm, and ${\kappa_{\text{min}}}({\rho_t})$ sets the smallest eigenvalue of the evolved state $\rho_t$. It is worthwhile noting that the right-hand side of Eq.~\eqref{eq:0000000000005} depends on the trace speed ${{\left\| {d}{{\rho}_t}/dt \right\|}_1}$, also known as Schatten speed~\cite{PhysRevA.97.022109}. Next, by combining Eqs.~\eqref{eq:0000000000004} and~\eqref{eq:0000000000005}, one gets that the rate of change of the UQE fulfills the upper bound
\begin{equation}
\label{eq:0000000000006}
\left| \frac{d}{dt}{{\text{E}}_{\alpha,\mu}}({\rho_t}) \right| \leq {h_{\alpha}}[{\kappa_{\text{min}}}({\rho_t})]{{\left\| \frac{d{{\rho}_t}}{dt} \right\|}_1} ~,
\end{equation}
where we define the auxiliary function
\begin{equation}
\label{eq:0000000000007}
{h_{\alpha}}[x] := \frac{\alpha}{1 - \alpha}\left(1 - \alpha + x \right) {x^{\alpha - 2}} ~.
\end{equation}
To obtain an upper bound on UQE for the general nonunitary dynamics, we integrate Eq.~\eqref{eq:0000000000006} over the interval $t \in [0 , \tau]$, and use the fact that $\left|{\int} \, dt \, g(t)\right| \leq {\int} \, dt \, |g(t)|$, which implies that
\begin{equation}
\label{eq:0000000000008}
\left| {{\text{E}}_{\alpha,\mu}}({\rho_{\tau}}) - {{\text{E}}_{\alpha,\mu}}({\rho_0})  \right| \leq {\int_0^{\tau}} dt\, {h_{\alpha}}[{\kappa_{\text{min}}}({\rho_t})] {{\left\| \frac{d{{\rho}_t}}{dt} \right\|}_1} ~.
\end{equation}

It is noteworthy that Eq.~\eqref{eq:0000000000008} is the first main result of this article. We see that the right-hand side of Eq.~\eqref{eq:0000000000008} depends on both the trace speed and the smallest eigenvalue of the evolved state. The Schatten norm $\|{d}{\rho_t}/{dt}\|_1$ cha\-rac\-te\-rizes the quantum speed that is induced by the ge\-ne\-ral physical process. In turn, such quantity can be explicitly evaluated by specifying the nonunitary evolution $\rho_t = {\mathcal{E}_t}(\rho_0)$, thus being rewritten in terms of the operators that generate the nonunitary dynamics. We find that the weight function ${h_{\alpha}}[{\kappa_{\text{min}}}({\rho_t})]$ depends on the smallest eigenvalue ${\kappa_{\text{min}}}({\rho_t})$ of the evolved density matrix, also being labeled by the parameter $\alpha$ that sets the UQE. We note that the left-hand side of Eq.~\eqref{eq:0000000000008} signals how far the initial and final states are by means of the absolute difference of UQEs for both the states. In turn, such a quantity is upper bounded by average speed, i.e., the Schatten 1-norm of the rate of change of the instantaneous state of the quantum system. It is worth mentioning that the bound requires low computational cost since its evaluation requires the smallest eigenvalue of the evolved state. For example, this can be useful in the study of UQE for higher dimensional systems, e.g., quantum many-body models, in which evaluating the full spectrum of the density matrix can be a formidable computational task.

In order to investigate the tightness of the bound on UQE in Eq.~\eqref{eq:0000000000008}, we introduce the figure of merit as follows
\begin{equation}
\label{eq:0000000000009}
{\delta_{\alpha,\mu}}(\tau) := 1 - \frac{\left| {\text{E}_{\alpha,\mu}}({\rho_{\tau}}) - {\text{E}_{\alpha,\mu}}({\rho_{0}}) \right|}{ {\int_0^{\tau}} dt\, {h_{\alpha}}[{\kappa_{\text{min}}}({\rho_t})] {{\left\| {d{{\rho}_t}}/{dt} \right\|}_1}} ~.
\end{equation}
Overall, the smaller the relative error in Eq.~\eqref{eq:0000000000009}, the tighter the bound on UQE in Eq.~\eqref{eq:0000000000008}. For our purposes, throughout the paper we will focus on the normalized relative error ${\widetilde{\delta}_{\alpha,\mu}}(\tau)$, with $\widetilde{x} := (x - \min(x))/(\max(x) - \min(x))$, noting that $0 \leq {\widetilde{\delta}_{\alpha,\mu}}(\tau) \leq 1$. In Secs.~\ref{sec:00000000005} and~\ref{sec:00000000011}, we will discuss in detail the relevance of this quantity. We emphasize that the tightness of the bound in Eq.~\eqref{eq:0000000000008} has a different meaning from the geometric perspective discussed in Refs.~\cite{PhysRevLett.65.1697,2013_PhysRevLett_110_050402,PhysRevX.6.021031,PhysRevLett.123.180403,PhysRevLett.127.100404}, for example. In those cases, given the evolution between initial and final states, the bound on QSL is saturated when the dynamical evolution coincides with the length of the geodesic path that connects the two states. Here, the tightness of Eq.~\eqref{eq:0000000000008} is assigned by the relative error in Eq.~\eqref{eq:0000000000009}, which in turn quantifies the deviation of the absolute difference of UQEs with respect to the average quantum speed. Therefore, the bound saturates when the rate of change of the unified entropy coincides with the product between the weight function and quantum speed induced by the nonunitary dynamics. In the following, starting from Eq.~\eqref{eq:0000000000008}, we present a family of speed limits for nonunitary processes that are related to the quantum unified entropy.


\section{UQE and Quantum Speed Limits}
\label{sec:00000000005}

How fast does a quantum system evolves under a given nonunitary dynamics? In this section we derive a two-parametric class of speed limits rooted on the UQE for quantum states evolving nonunitarily. In essence, the QSL denotes the minimum time of evolution for quantum states undergoing a certain dynamics. For closed quantum systems, the QSL time for orthogonal states $|{\psi_0}\rangle$ and $|{\psi_{\tau}}\rangle$ is given by ${\tau_{\text{QSL}}} = \text{max}\{ \hbar \pi/(2\Delta{E}), \hbar \pi/(2{\langle{\psi_0}|{H}|{\psi_0}\rangle}) \}$, where ${(\Delta{E})^2} = \langle{\psi_0}|{H^2}|{\psi_0}\rangle - {\langle{\psi_0}|{H}|{\psi_0}\rangle^2}$ is the variance of the time-independent Hamiltonian of the system~\cite{1945_JPhysURSS_9_249,1992_PhysicaD_120_188,PhysRevLett.103.160502}. In this regard, we point out that Ref.~\cite{PhysRevLett.127.100404} presented a unified QSL bound by means of the changing rate of phase of the quantum system, which in turn is based on the transition speed of states and the accumulation phase of the quantum system. Noteworthy, QSLs have been addressed in different scenarios, and find applications ranging from the dynamics of either closed and open quantum systems~\cite{2013_PhysRevLett_110_050402,PhysRevLett.110.050403,2013_PhysRevLett_111_010402,PhysRevA.97.052125,PhysRevLett.123.180403,PhysRevA.103.022210,arXiv:2110.13193}, to quantum many-body systems~\cite{PhysRevX.9.011034,PhysRevResearch.2.023299,PhysRevA.102.042606,PhysRevResearch.2.032020,arXiv:2006.14523,PhysRevLett.124.110601,PhysRevLett.126.180603,PhysRevX.11.011035,PhysRevX.12.011038}, also including the discussion of quantum-classical limits~\cite{PhysRevLett.120.070401,PhysRevLett.120.070402,arXiv:2112.09716}, quantum thermodynamics~\cite{PhysRevA.102.042606,PhysRevE.103.032105,PhysRevA.104.052223,arXiv:2203.12421}, and non-Hermitian systems~\cite{ChinPhysB_29_030304,PhysRevA.104.052620}. Interestingly, QSLs have been also applied to the study of the dynamics of cosmological coherent states in the context of loop quantum gravity~\cite{ClassQuantumGrav_39_12LT01}. We refer to Ref.~\cite{Deffner_2017} for a recent topic review on quantum speed limits and its applications.

Here we put forward this discussion and address a fami\-ly of QSLs based on unified entropies. In detail, by considering the UQE as a useful information-theoretic quantifier, we present a lower bound on the time of evolution for nonunitary physical processes. Starting from Eq.~\eqref{eq:0000000000008}, the time required for an arbitrary nonunitary evolution driving the quantum system from $\rho_0$ to $\rho_{\tau}$ is lower bounded as $\tau \geq {{\tau}^{\text{QSL}}_{\alpha,\mu}}$, with the QSL time given by
\begin{equation}
\label{eq:00000000000011}
{{\tau}^{\text{QSL}}_{\alpha,\mu}} := \frac{\left| {{\text{E}}_{\alpha,\mu}}({\rho_{\tau}}) - {{\text{E}}_{\alpha,\mu}}({\rho_0})  \right|}{ {\langle\langle {h_{\alpha}}[{\kappa_{\text{min}}}({\rho_t})]\, {{\left\| {d{{\rho}_t}}/{dt} \right\|}_1} \rangle\rangle_{\tau}}}  ~,
\end{equation}
where ${\langle\langle g(t) \rangle\rangle_{\tau}} := \frac{1}{\tau}{\int_0^{\tau}}\, dt \, g(t)$ stands for the time ave\-ra\-ge, and $0 < \alpha < 1$ and $0 < \mu < 1$. We stress that Eq.~\eqref{eq:00000000000011} is the second main result of the paper. The QSL in Eq.~\eqref{eq:00000000000011} is a time-dependent function, which comes from the fact that we are distinguishing two arbitrary states $\rho_0$ and $\rho_{\tau}$. Indeed, we note that the absolute value of the difference of UQEs in Eq.~\eqref{eq:00000000000011} captures the distinguishability of initial and final states, somehow assigning a geometric interpretation in terms of a distance between these states. Note that UQE satisfies the so-called Lipschitz continuity $\left| {{\text{E}}_{\alpha,\mu}}({\rho_{\tau}}) - {{\text{E}}_{\alpha,\mu}}({\rho_0})  \right| \leq {(\alpha(\alpha - 1))^{-1}}\|{\rho_{\tau}} - {\rho_0}\|_1$ [see Sec.~\ref{sec:00000000002}]. We see that the absolute difference of UQEs is upper bounded by the trace distance of initial and final states~\cite{Nielsen_Chuang_infor_geom}. The absolute difference of UQEs must be smaller than the trace distance, which in turn attributes the notion of how far apart two neighboring states are in the space of quantum states. Note that the absolute difference of UQEs also signals the sensitivity of the figure of merit respective to the dynamics generated by the nonunitary evolution.

In addition, the QSL depends on the Schatten speed, which in turn depicts the quantum speed induced by the dynamics respective to changes in time over the interval $t \in [0,\tau]$. This means that $\tau_{\text{QSL}}$ is related to the dynamics of the eigenstates of the generators that govern the nonunitary evolution of the system~\cite{Deffner_2017}. The QSL time is inversely proportional to the average speed, given the weight function depending on the smallest eigenvalue of the state. We point out that the tightness of the QSL in Eq.~\eqref{eq:00000000000011} is related to the tightness of the bound in Eq.~\eqref{eq:0000000000008}. Hence, the relative error in Eq.~\eqref{eq:0000000000009} stands as a useful figure of merit to infer the tightness of the QSL bound presented in Eq.~\eqref{eq:00000000000011}.

Next, we comment on the behavior of the QSL in Eq.~\eqref{eq:00000000000011} in the limiting cases of R\'{e}nyi and Tsallis entropies. On the one hand, UQE reduces to the R\'{e}nyi entropy for $\mu \rightarrow 0$, and one gets ${{\tau}^{\text{QSL}}_{\alpha,0}} = {\left| {{\text{R}}_{\alpha}}({\rho_{\tau}}) - {{\text{R}}_{\alpha}}({\rho_0})  \right|}/{ {\langle\langle {h_{\alpha}}[{\kappa_{\text{min}}}({\rho_t})]\, {{\left\| {d{{\rho}_t}}/{dt} \right\|}_1} \rangle\rangle_{\tau}}}$. In turn, by taking the limiting case $\alpha \rightarrow 1$, the R\'{e}nyi entropy collapses into the von Neumann entropy, while ${\lim_{\alpha \rightarrow 1}}\,{h_{\alpha}}[{\kappa_{\text{min}}}({\rho_t})] \rightarrow \infty$ [see Eq.~\eqref{eq:0000000000007}], and one concludes that ${\lim_{\alpha \rightarrow 1}} \, {\tau^{\text{QSL}}_{\alpha,0}} \approx 0$. On the other hand, for $\mu \rightarrow 1$, UQE reduces further to the Tsallis entropy, and the QSL becomes ${{\tau}^{\text{QSL}}_{\alpha,1}} = {\left| {{\text{H}}_{\alpha}}({\rho_{\tau}}) - {{\text{H}}_{\alpha}}({\rho_0})  \right|}/{ {\langle\langle {h_{\alpha}}[{\kappa_{\text{min}}}({\rho_t})]\, {{\left\| {d{{\rho}_t}}/{dt} \right\|}_1} \rangle\rangle_{\tau}}}$. In the limit $\alpha \rightarrow 1$, Tsallis entropy also recovers the von Neumann entropy, and we expect that ${\lim_{\alpha \rightarrow 1}} {\tau^{\text{QSL}}_{\alpha,1}} \approx 0$ approaches zero in the same way as in the previous case since ${\lim_{\alpha \rightarrow 1}}\,{h_{\alpha}}[{\kappa_{\text{min}}}({\rho_t})] \rightarrow \infty$. In both the aforementioned limiting cases, one gets the lower bound $\tau \geq 0$. This means that, for $\mu \rightarrow 0$ (or $\mu \rightarrow 1$) and $\alpha \rightarrow 1$, these bounds become insensitive to the nonunitary dynamic properties of the quantum system, thus implying a trivial QSL.

In the following we will specialize the results in Eqs.~\eqref{eq:0000000000008},~\eqref{eq:0000000000009}, and~\eqref{eq:00000000000011} in view of two paradigmatic nonunitary evolutions, thus specifying the Schatten speed for each case. The first one is given by quantum channels, i.e., completely positive dynamical maps. The second case addresses the nonunitary dynamics of dissipative quantum systems that can be described by effective non-Hermitian Hamiltonians. To illustrate our findings, we investigate the dynamics of a single-qubit state undergoing each of these evolutions.


\begin{figure*}[!t]
\begin{center}
\includegraphics[scale=0.8]{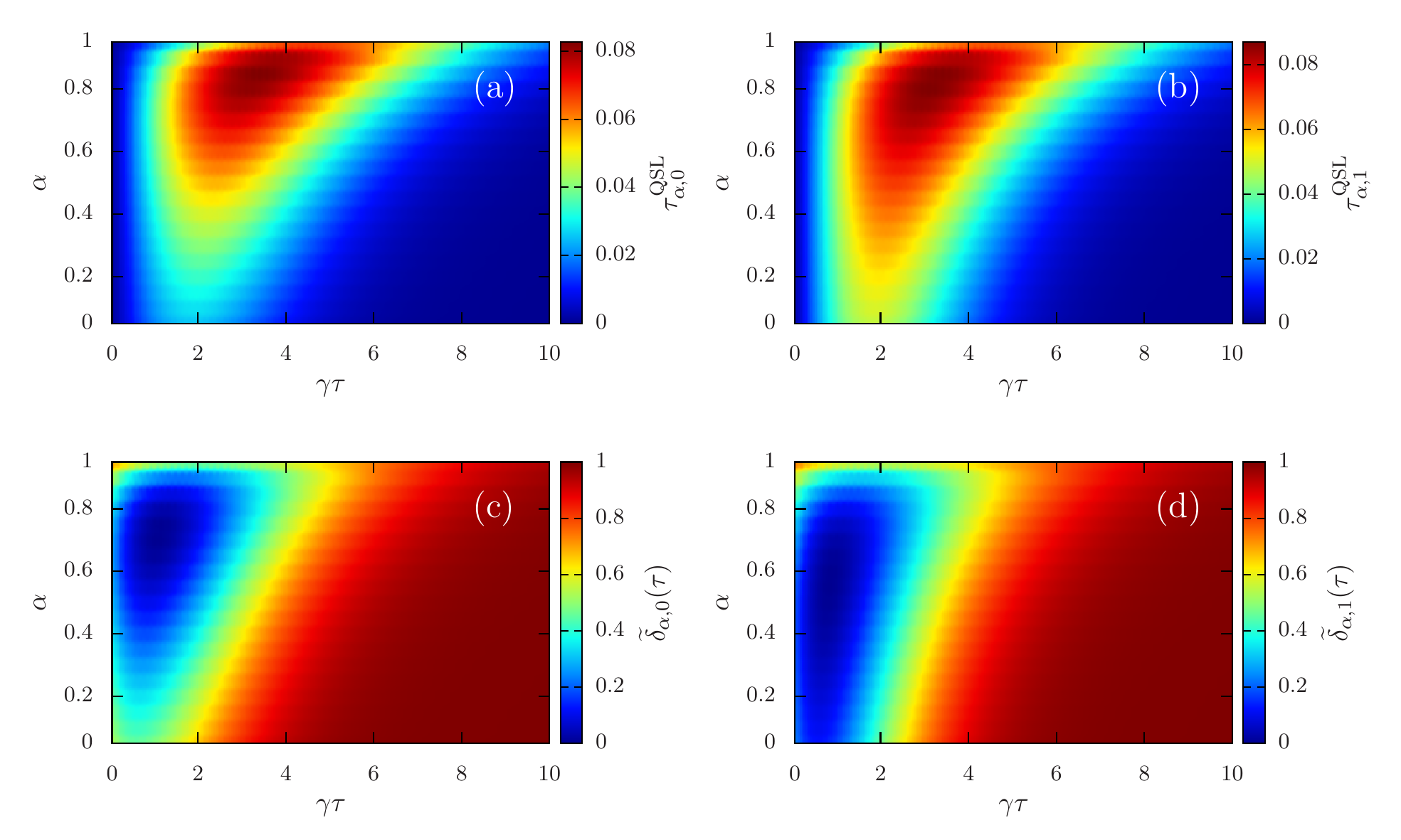}
\caption{(Color online) Density plot of QSL time $\tau^{\text{QSL}}_{\alpha,\mu}$ [see Eq.~\eqref{eq:00000000000017}] and the normalized relative error ${\widetilde{\delta}_{\alpha,\mu}}(\tau)$ [see Eq.~\eqref{eq:00000000000015}], as a function of the dimensionless parameter $\gamma\tau$, for a single-qubit state evolving under the amplitude damping channel. Here we choose the initial state ${\rho_0} = (1/2)(\mathbb{I} + \vec{r}\cdot\vec{\sigma})$ with $\{r,\theta\} = \{1/2,\pi/4,\pi/4\}$.}
\label{fig:FIG02}
\end{center}
\end{figure*}

\subsection{Quantum channels}
\label{sec:00000000007}

We shall begin considering the dynamical evolution generated by the completely-positive and trace-preserving (CPTP) quantum operation, ${{\mathcal{E}}_t}(\bullet) = {\sum_{\ell}}\,{K_{\ell}} \bullet {K^{\dagger}_{\ell}}$, with $\{{K_{\ell}}\}_{\ell = 1,\ldots,q}$ being the set of time-dependent Kraus ope\-ra\-tors fulfilling ${\sum_{\ell}}\, {K^{\dagger}_{\ell}}{K_{\ell}} = \mathbb{I}$~\cite{Nielsen_Chuang_infor_geom}. In this case, we obtain that the trace speed is upper bounded as
\begin{equation}
\label{eq:00000000000012}
{{\left\| \frac{d {{\rho}_t}}{dt} \right\|}_1} \leq 2 \, {\sum_{\ell}} \, {{\|{K_{\ell}}{\rho_0}{\dot{K}_{\ell}^{\dagger}}\|}_1} ~,
\end{equation}
where we have used the triangular inequality ${{\| {\sum_{\ell}} \, {{\mathcal{O}}_{\ell}} \|}_1} \leq {\sum_{\ell}} {{\| {{\mathcal{O}}_{\ell}} \|}_1}$, and the unitary invariance of the Schatten norm as ${{\| {{\mathcal{O}}^{\dagger}} \|}_1} = {{\| {{\mathcal{O}}} \|}_1}$. Hence, Eq.~\eqref{eq:0000000000008} becomes
\begin{align}
\label{eq:00000000000013}
&\left| {{\text{E}}_{\alpha,\mu}}({\rho_{\tau}}) - {{\text{E}}_{\alpha,\mu}}({\rho_0})  \right| \nonumber\\
&\leq  2 \tau \,{\sum_{\ell}} {\langle\langle {h_{\alpha}}[{\kappa_{\text{min}}}({\rho_t})]\, {{\|{K_{\ell}}{\rho_0}{\dot{K}_{\ell}^{\dagger}}\|}_1} \rangle\rangle_{\tau}} ~,
\end{align}
which in turn allows us to write the relative error
\begin{equation}
\label{eq:00000000000015}
{\delta_{\alpha,\mu}}(\tau) = 1 - \frac{\left| {{\text{E}}_{\alpha,\mu}}({\rho_{\tau}}) - {{\text{E}}_{\alpha,\mu}}({\rho_0})  \right|}{2 \,{\sum_{\ell}} {\int_0^{\tau}} dt \, {h_{\alpha}}[{\kappa_{\text{min}}}({\rho_t})]\, {{\|{K_{\ell}}{\rho_0}{\dot{K}_{\ell}^{\dagger}}\|}_1}} ~.
\end{equation}
In addition, by substituting Eq.~\eqref{eq:00000000000012} into Eq.~\eqref{eq:00000000000011}, one gets the QSL time as follows:
\begin{equation}
\label{eq:00000000000014}
{\tau^{\text{QSL}}_{\alpha,\mu}} = \frac{ \left| {{\text{E}}_{\alpha,\mu}}({\rho_{\tau}}) - {{\text{E}}_{\alpha,\mu}}({\rho_0})  \right|}{ 2\, {\sum_{\ell}} {\langle\langle {h_{\alpha}}[{\kappa_{\text{min}}}({\rho_t})]\, {{\|{K_{\ell}}{\rho_0}{\dot{K}_{\ell}^{\dagger}}\|}_1} \rangle\rangle_{\tau}}} ~.
\end{equation}
It is noteworthy that Eq.~\eqref{eq:00000000000014} provides a lower bound on the time of evolution between states ${{\rho}_0}$ and ${{\rho}_{\tau}}$ for a system evol\-ving under a CPTP map, as a function of the absolute value of the UQE. The bound depends on the set of Kraus operators that characterizes the evolution, and the smallest eigenvalue of the evolved state. In the following, we will discuss Eqs.~\eqref{eq:00000000000014} and~\eqref{eq:00000000000015} focusing on a single-qubit state.


\subsubsection{Amplitude damping channel}
\label{sec:00000000008}

Here we consider a two-level system undergoing the noisy evolution modeled by an amplitude damping channel, which is given by the following Kraus operators $K_0 = |0\rangle\langle{0}| + {e^{-\gamma t/2}}\,|{1}\rangle\langle{1}|$, and $K_1 = \sqrt{1 - {e^{-\gamma t}}}\, |0\rangle\langle{1}|$. Here $\{ |0\rangle,|{1}\rangle\}$ stand for the computational basis states in the complex two-dimensional vector space ${\mathbb{C}^2}$, while $\gamma^{-1}$ defines the cha\-rac\-te\-ris\-tic time of the dissipative process~\cite{PhysRevA.102.012401}. The system is initialized at the single-qubit state $\rho_0 = (1/2)(\mathbb{I} + \vec{r}\cdot\vec{\sigma})$, with $\mathbb{I}$ the $2\times 2$ identity matrix, where $\vec{r} = \{r\sin\theta\cos\phi, r\sin\theta\sin\phi , r\cos\theta \}$ is the Bloch vector, with $r \in [0,1]$, $\theta \in [0,\pi]$ and $\phi \in [0,2\pi [$, while $\vec{\sigma} = \{{\sigma_x},{\sigma_y},{\sigma_z}\}$ is the vector of Pauli matrices. The evolved state is given by ${\rho_t} = {\sum_{\ell = 1,2}}\,{K_{\ell}}{\rho_0}{K_{\ell}^{\dagger}}$, and its $\alpha$-purity reads as ${f_{\alpha}}({\rho_t}) = {({\kappa_{\text{min}}}({\rho_t}))^{\alpha}} + {(1 - {\kappa_{\text{min}}}({\rho_t}))^{\alpha}}$, with the smallest eigenvalue ${\kappa_{\text{min}}}({\rho_t}) = (1/2)(1 - \sqrt{1 - {e^{-\gamma t}}{\xi_t}} \,)$, and the auxiliary function ${\xi_t} = 1 - {r^2} + (1 - {e^{-\gamma t}}){(1 - r\cos\theta)^2}$. In particular, for $t = 0$, $\alpha$-purity reduces to ${f_{\alpha}}({\rho_0}) = {2^{-\alpha}}({(1 - r)^{\alpha}} + {(1 + r)^{\alpha}})$. The QSL time respective to the nonunitary evolution between states $\rho_0$ and $\rho_{\tau}$ is given by
\begin{widetext}
\begin{equation}
\label{eq:00000000000017}
{\tau^{\text{QSL}}_{\alpha,\mu}} = \frac{\left| {\left( {\kappa^{\alpha}_{\text{min}}}({\rho_{\tau}}) + {\left( 1 - {\kappa_{\text{min}}}({\rho_{\tau}})\right)^{\alpha}}\right)^{\mu}}  - {2^{-\alpha\mu}}\left({(1 - r)^{\alpha}} + {(1 + r)^{\alpha}}\right)^{\mu} \right|}{\frac{1}{2}\gamma \alpha\mu \, \frac{1}{\tau}{\int_0^{\tau}} \, dt \, {e^{-\gamma t}}(1 - \alpha + {\kappa_{\text{min}}}({\rho_t})) \, {{\kappa^{\alpha - 2}_{\text{min}}}({\rho_t})}\left(1 - r\cos\theta + \sqrt{{(1 - r\cos\theta)^2} + {e^{\gamma t}}{r^2}{\sin^2}\theta}\, \right) } ~.
\end{equation}
\end{widetext}

In Fig.~\ref{fig:FIG02}, we show plots of the QSL time and the re\-la\-tive error for the for the initial single-qubit state parametrized as $\{r,\theta,\phi\} = \{1/2,\pi/4,\pi/4\}$. In Figs.~\ref{fig:FIG02}(a) and~\ref{fig:FIG02}(b), we show the plots of the QSL in Eq.~\eqref{eq:00000000000017} for R\'{e}nyi ($\mu = 0$) and Tsallis entropies ($\mu = 1$), respectively. We see that, regardless $\alpha \in (0,1)$, both quantities $\tau^{\text{QSL}}_{\alpha,1}$ and $\tau^{\text{QSL}}_{\alpha,0}$ exhibit nonzero values at earlier times, and approach zero at later times of the nonunitary dynamics. Overall, the two QSLs show slightly different quantitative behaviors, but the same qualitative features. 

In Figs.~\ref{fig:FIG02}(c) and~\ref{fig:FIG02}(d), we plot the normalized relative error ${\widetilde{\delta}_{\alpha,\mu}}(\tau)$ for the R\'{e}nyi and Tsallis entropies, respectively [see Eq.~\eqref{eq:00000000000015}]. We find that, for all $0 < \alpha < 1$, such a quantity approaches small values around $0 \lesssim \gamma\tau \lesssim 3$, while it becomes close to the unit at later times of the dynamics. We see that ${\lim_{\tau \rightarrow \infty}}\, {\widetilde{\delta}_{\alpha,\mu}}(\tau) \approx 1$ for all $\alpha \in (0,1)$, and the QSL $\tau^{\text{QSL}}_{\alpha,\mu} \approx 0$ approaches zero [see Figs.~\ref{fig:FIG02}(a) and~\ref{fig:FIG02}(b)], thus implying that $\tau \gtrsim 0$ as time increases. This means that, the larger the time, the looser is the QSL bound in Eq.~\eqref{eq:00000000000017}. To see this, we first remind that the amplitude damping channel shrinks the Bloch sphere towards the north pole. In detail, for $\tau \gg {\gamma^{-1}}$, the evolved state of the two-level system approaches the incoherent pure state $|{0}\rangle\langle{0}|$, which in turn has a vanishing UQE, and a zero-valued smallest eigenvalue. In this case, we expect the smallest eigenvalue of $\rho_{\tau}$ to exhibit asymptotic behavior ${\kappa_{\text{min}}}({\rho_{\tau}}) \approx {\kappa_{\text{min}}}(|{0}\rangle\langle{0}|) \approx 0$ at later times, which implies that the function ${h_{\alpha}}[{\kappa_{\text{min}}}({\rho_{\tau}})] \approx {h_{\alpha}}[{\kappa_{\text{min}}}(|{0}\rangle\langle{0}|)]$ will assume larger values, for all $\alpha \in (0,1)$ [see Eq.~\eqref{eq:0000000000007}]. In this case, one arrives at the trivial bound $\tau \gtrsim 0$.


\subsection{Non-Hermitian evolution}
\label{sec:00000000009}

\begin{figure*}[!t]
\begin{center}
\includegraphics[scale=0.65]{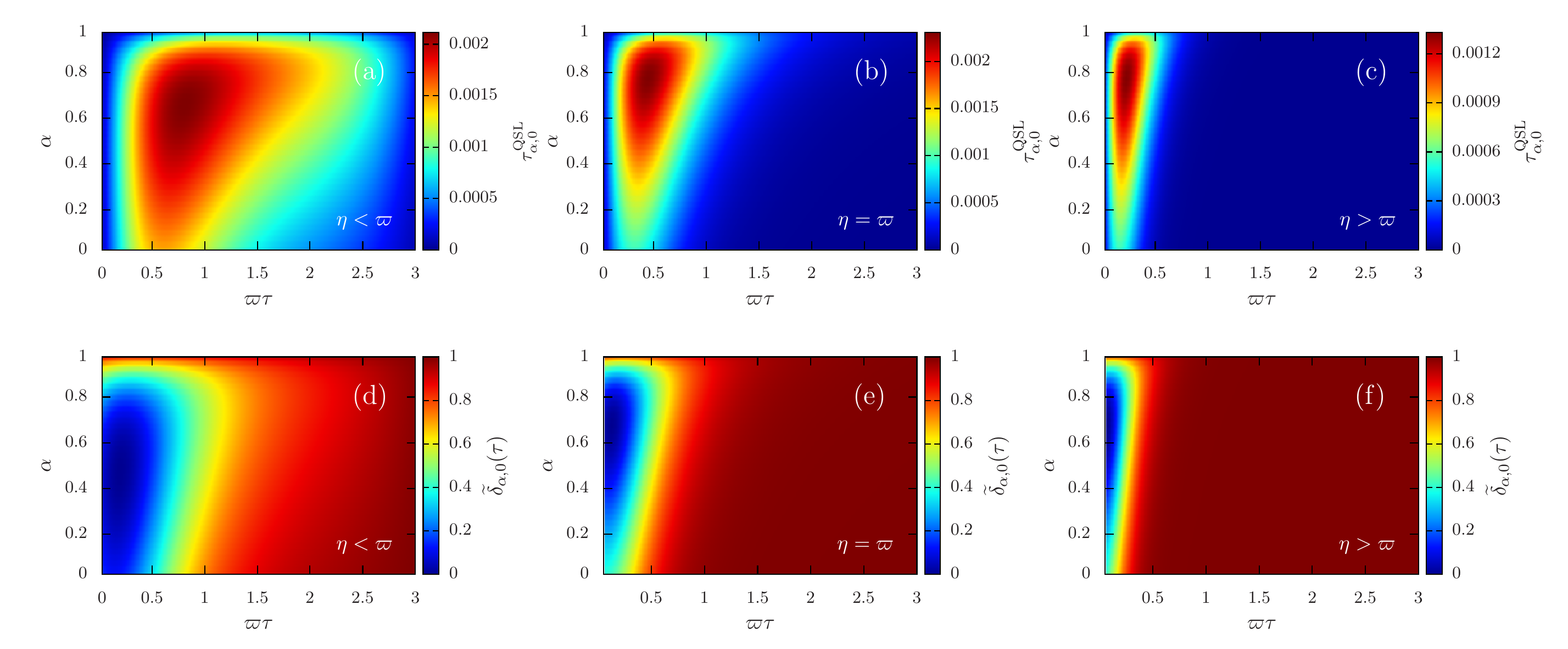}
\caption{(Color online) Plot of the QSL time $\tau^{\,\text{QSL}}_{\alpha,0}$ [see Eq.~\eqref{eq:00000000000019}], and the relative error ${\widetilde{\delta}_{\alpha,0}}(\tau)$ [see Eq.~\eqref{eq:00000000000015}], for the non-Hermitian two-level system described by the Hamiltonian $\widetilde{H} = \varpi{\sigma_x} + i{\eta}{\sigma_z}$. The system is initialized at the single-qubit state ${\rho_0} = (1/2)(\mathbb{I} + \vec{r}\cdot\vec{\sigma})$, with $\{r,\theta,\phi\} = \{0.5,\pi/4,\pi/4\}$. Here we set the cases $\eta/\varpi = 0.5$ [panels (a), (d)], $\eta/\varpi = 1$ [panels (b), (e)], and $\eta/\varpi = 2$ [panels (c), (f)].}
\label{fig:FIG03}
\end{center}
\end{figure*}

Dissipative systems whose dynamics can be modeled by effective non-Hermitian Hamiltonians have triggered considerable efforts in theoretical and experimental research~\cite{PhysRevLett.80.5243,Ramy2018,PhysRevLett.123.090603,Longhi2019.6,ProgTheorExpPhys_12_2020,PhysRevA.101.062112,PhysRevB.104.155141,arXiv:2111.04378,arXiv:2201.05367}. We note, for instance, that Refs.~\cite{PhysRevLett.110.050403,PhysRevLett.123.180403,PhysRevLett.127.100404} have addressed the study of the quantum speed limit for the dynamics of quantum systems described by non-Hermitian Hamiltonians. In the following, by using Eqs.~\eqref{eq:0000000000008},~\eqref{eq:0000000000009}, and~\eqref{eq:00000000000011}, we investigate the interplay between UQE and QSL for a quantum system undergoing the nonunitary evolution ${{\mathcal{E}}_t}(\bullet) = {e^{-i t \widetilde{H}}} \bullet\,  {e^{+ i t {\widetilde{H}^{\dagger}} }}$ that is generated by a general effective non-Hermitian Hamiltonian $\widetilde{H} = {H} + i{\Gamma}$, where $H = (1/2)(\widetilde{H} + {\widetilde{H}^{\dagger}})$ and ${\Gamma} = (1/2i)(\widetilde{H} - {\widetilde{H}^{\dagger}})$ are Hermitian operators. The normalized time-dependent quantum state ${\rho_t} = {{\mathcal{E}}_t}({\rho_0})/\text{Tr}({{\mathcal{E}}_t}({\rho_0}))$ fulfills the equation of motion ${d{\rho_t}}/{dt} = -i [{H},{\rho_t}] + \{{\Gamma},{\rho_t} \} - 2{\langle{\Gamma}\rangle_t}\,{\rho_t}$, where ${\langle\bullet\rangle_t} = \text{Tr}(\bullet {\rho_t})$ stands for the expectation value at time $t \geq 0$~\cite{PhysRevA.42.1467,doi:10.1142,EPJD_253_69_2015}. In this case, the Schatten speed is bounded from above as
\begin{equation}
\label{eq:00000000000018}
{\left\| \frac{d{\rho_t}}{dt} \right\|_1} \leq 2({\| H \|_{\infty}} + {\| \Gamma \|_1} + {\| \Gamma \|_{\infty}}) ~, 
\end{equation}
where we have invoked the following set of inequa\-li\-ties ${\|{A_1} + {A_2} \|_1} \leq {\|{A_1}\|_1} + {\|{A_2}\|_1}$, ${\| [{A_1},{A_2}] \|_1} \leq 2{\|{A_1}\|_{\infty}}{\|{A_2}\|_1}$, ${\| {A_1}{A_2} \|_1} \leq {\|{A_1}\|_1}{\|{A_2}\|_1}$, and $|\text{Tr}({A_1}{A_2})| \leq {\|{A_1}\|_{\infty}}{\|{A_2}\|_1}$~\cite{Bathia_Rajendra,FONG20111193,CHENG2015409}. In addition, we have used the identity ${\|{\rho_t} \|_1} = 1$, which comes from the fact that ${\rho_t} \in \mathcal{S}$ [see Sec.~\ref{sec:00000000002}]. Therefore, by combining Eqs.~\eqref{eq:00000000000011} and~\eqref{eq:00000000000018}, the QSL time for the non-Hermitian dynamics thus yields
\begin{equation}
\label{eq:00000000000019}
{\tau^{\text{QSL}}_{\alpha,\mu}} = \frac{ \left| {{\text{E}}_{\alpha,\mu}}({\rho_{\tau}}) - {{\text{E}}_{\alpha,\mu}}({\rho_0}) \right|}{ 2\, {\left\langle\left\langle {h_{\alpha}}({\rho_t})({\| H \|_{\infty}} + {\| \Gamma \|_1} + {\| \Gamma \|_{\infty}}) \right\rangle\right\rangle_{\tau}} } ~,
\end{equation}
with ${\rho_{\tau}} = {{\mathcal{E}}_{\tau}}({\rho_0})/\text{Tr}({{\mathcal{E}}_{\tau}}({\rho_0}))$, and we introduce the relative error
\begin{equation}
\label{eq:00000000000020}
{\delta_{\alpha,\mu}}(\tau) = 1 - \frac{\left| {{\text{E}}_{\alpha,\mu}}({\rho_{\tau}}) - {{\text{E}}_{\alpha,\mu}}({\rho_0}) \right|}{ 2\, {\int_0^{\tau}} dt \, {h_{\alpha}}({\rho_t})({\| H \|_{\infty}} + {\| \Gamma \|_1} + {\| \Gamma \|_{\infty}})  } ~.
\end{equation}
We see that, apart from the smallest eigenvalue of the evolved state, the lower bound in Eq.~\eqref{eq:00000000000019} depends on operators $H$ and $\Gamma$, and the unified entropy of both the initial and final states of the system. In parti\-cu\-lar, note that the QSL in Eq.~\eqref{eq:00000000000019} vanishes in the Hermitian limit $\widetilde{H} = H$ ($\Gamma = 0$), regardless $0 < \alpha < 1$ and $0 < \mu < 1$. This result comes from the fact that UQE remains invariant when the system undergoes a fully unitary dynamics, i.e., ${[{{\text{E}}_{\alpha,\mu}}({\rho_t})]_{\Gamma = 0}} = {{\text{E}}_{\alpha,\mu}}({e^{- i t H}} {\rho_0} \, {e^{i t H}})  = {{\text{E}}_{\alpha,\mu}}({\rho_0})$ for all $t \in [0,\tau]$, thus implying the trivial bound $\tau \geq 0$. Note that the result in Eq.~\eqref{eq:00000000000019} differs from the QSL time discussed in Ref.~\cite{PhysRevLett.110.050403}, the latter depending on the relative purity of the initial and final states of the system, and being inversely proportional to the variance of the real and imaginary parts of the non-Hermitian Hamiltonian. Furthermore, we shall mention that the bound in Ref.~\cite{PhysRevLett.123.180403} stands as a general result, which in turn exhibits a clear geometric interpretation in terms of the geometric phase of the quantum system, also being tighter than the Mandelstam-Tamm and Margolus-Levitin bounds in some cases.


\subsubsection{$\mathcal{P}\mathcal{T}$-symmetric non-Hermitian Hamiltonian}
\label{sec:00000000010}

Here, we specialize the results of Sec.~\ref{sec:00000000009} to the parity-time-reversal ($\mathcal{P}\mathcal{T}$) symmetric non-Hermitian Hamiltonian $\widetilde{H} = \varpi{\sigma_x} + i\eta{\sigma_z}$, where $\varpi \in \mathbb{R}$ plays the role of a coupling strength, and $\eta \in \mathbb{R}$ denotes a dissipation rate~\cite{10.1093ptepptaa181}. It is noteworthy that this system exhibits three phases: (i) unbroken $\mathcal{P}\mathcal{T}$ symmetry-preserving phase ($\eta/\varpi < 1$) with real eigen\-va\-lues $\pm \sqrt{{\varpi^2} - {\eta^2}}$; (ii) gapless, critical phase, at the exceptional point $\eta = \varpi$, where the spectrum coalesces; (iii) $\mathcal{P}\mathcal{T}$ symmetry-broken phase ($\eta/\varpi > 1$), in which the eigenvalues become purely imaginary $\pm i\, \sqrt{{\eta^2} - {\varpi^2}}$. Recently, this system has been experimentally rea\-li\-zed in dissipative Floquet systems~\cite{NatComm_10_855_2019}, trapped ion setups~\cite{PhysRevLett.126.083604}, and with nuclear spins~\cite{npj_QuInf_6_1_2020}. The system is initialized at the state $\rho_0 = (1/2)(\mathbb{I} + \vec{r}\cdot\vec{\sigma})$, with $\vec{r} = \{r\sin\theta\cos\phi, r\sin\theta\sin\phi , r\cos\theta \}$, $\vec{\sigma} = \{{\sigma_x},{\sigma_y},{\sigma_z}\}$, where $r \in [0,1]$, $\theta \in [0,\pi]$ and $\phi \in [0,2\pi [$. In Appendix~\ref{sec:00000000015}, we present details on the nonunitary dynamics of such single-qubit state under the non-Hermitian Hamiltonian.

In Fig.~\ref{fig:FIG03}, we plot the QSL time in Eq.~\eqref{eq:00000000000019} [see Figs.~\ref{fig:FIG03}(a)--\ref{fig:FIG03}(c)], and the relative error in Eq.~\eqref{eq:00000000000020} [see Figs.~\ref{fig:FIG03}(d)--\ref{fig:FIG03}(f)], as a function of the dimensionless parameter $\varpi\tau$, for the evolved state ${\rho_{\tau}} = {e^{-i \tau \widetilde{H}}} {\rho_0}\, {e^{+ i \tau {\widetilde{H}^{\dagger}} }}/\text{Tr}({e^{-i \tau \widetilde{H}}} {\rho_0}\, {e^{+ i \tau {\widetilde{H}^{\dagger}} }})$. Here we address the case of the R\'{e}nyi entropy ($\mu = 0$), and note that si\-mi\-lar results hold for Tsallis entropy ($\mu = 1$). Figures~\ref{fig:FIG03}(a) and~\ref{fig:FIG03}(d) show the QSL time and the re\-la\-tive error for the unbroken $\mathcal{P}\mathcal{T}$ symmetry-preserving phase $\eta/\varpi < 1$, with $\eta = 0.5\varpi$. Figure~\ref{fig:FIG03}(a) shows that, for $\alpha \in (0,1)$, the QSL time grows and exhibits nonzero values as $\varpi\tau$ increases, and smoothly approaches zero at later times. In Fig.~\ref{fig:FIG03}(d), we see that the relative error is small at earlier times, while approaches the unity as time increases. We point out that, the smaller the relative error, the tighter is the lower bound on UQE. 

In Figs.~\ref{fig:FIG03}(b) and~\ref{fig:FIG03}(e), we plot the QSL time and the relative error at the exceptional point $\eta = \varpi$. For $0 < \alpha < 1$, one verifies that the QSL time increases and reaches nonzero values for $0 \lesssim \varpi\tau \lesssim 1$, while it becomes zero at later times. In turn, the relative error indicates the bound in Eq.~\eqref{eq:00000000000019} becomes loose at later times. Next, Figs.~\ref{fig:FIG03}(c) and~\ref{fig:FIG03}(f) show the QSL time and the relative error for the case of $\mathcal{P}\mathcal{T}$ symmetry-broken phase $\eta/\varpi > 1$, with $\eta = 2\varpi$. In this case, for $0 < \alpha < 1$, we find that the QSL is nonzero for $0 \lesssim \varpi\tau \lesssim 0.5$ [see Fig~\ref{fig:FIG03}(c)], and becomes loose for $\varpi\tau \gtrsim 0.5$ as signaled by the relative error in Fig~\ref{fig:FIG03}(f). To summarize, note that the QSL displays nonzero values in a region in the $\varpi\tau$--$\alpha$ plane that decreases as the ratio $\eta/\varpi$ increases.

\begin{figure}[!t]
\begin{center}
\includegraphics[scale=0.95]{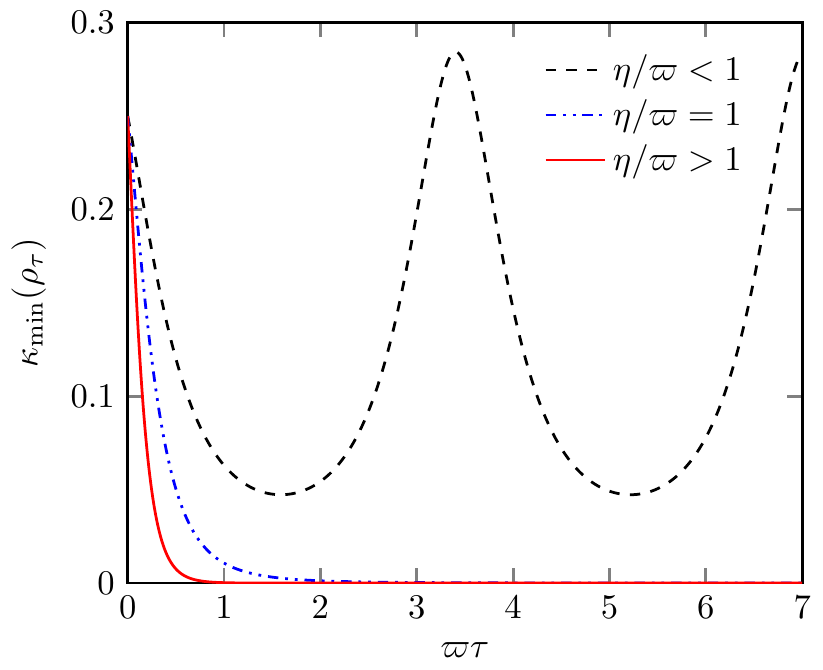}
\caption{(Color online) Plot of the smallest eigenvalue ${\kappa_{\text{min}}}({\rho_{\tau}})$ as a function of the dimensionless parameter $\varpi \tau$. Here ${\rho_{\tau}} = {e^{-i \tau \widetilde{H}}} {\rho_0}\, {e^{+ i \tau {\widetilde{H}^{\dagger}} }}/\text{Tr}({e^{-i \tau \widetilde{H}}} {\rho_0}\, {e^{+ i \tau {\widetilde{H}^{\dagger}} }})$ is the evolved state of the system, with $\widetilde{H} = \varpi{\sigma_x} + i{\eta}{\sigma_z}$ being a two-level non-Hermitian Hamiltonian. We set the initial single-qubit state ${\rho_0} = (1/2)(\mathbb{I} + \vec{r}\cdot\vec{\sigma})$, with $\{r,\theta,\phi\} = \{0.5,\pi/4,\pi/4\}$. Here we plot the cases $\eta/\varpi = 0.5$ (black dashed line), $\eta/\varpi = 1$ (blue dot dashed line), and $\eta/\varpi = 2$ (red solid line).}
\label{fig:FIG04}
\end{center}
\end{figure}

In the following, we comment on the behavior of the QSL and the relative error. In Fig.~\ref{fig:FIG04} we plot the smallest eigenvalue ${\kappa_{\text{min}}}({\rho_{\tau}})$ of the evolved state of the two-level system, as a function of $\varpi\tau$. On the one hand, for $\eta/\varpi < 1$ ($\mathcal{P}\mathcal{T}$ unbroken phase), ${\kappa_{\text{min}}}({\rho_{\tau}})$ is nonzero and periodically oscillates as a function of $\varpi\tau$, and the QSL exhibit nontrivial results [see Fig.~\ref{fig:FIG03}(a)]. On the other hand, at the exceptional point $\eta/\varpi = 1$, we find that ${\kappa_{\text{min}}}({\rho_{\tau}})$ takes nonzero values at earlier times, and suddenly vanishes for $\varpi\tau \gtrsim 2$. This means that the QSL time becomes loose at later times [see Fig.~\ref{fig:FIG03}(b)], which is witnessed by the relative error [see Fig.~\ref{fig:FIG03}(e)]. Note that this behavior persists in the $\mathcal{P}\mathcal{T}$ symmetry-broken phase with $\eta/\varpi > 1$, with the smallest eigenvalue decaying faster as the dissipative effect of the non-Hermitian dynamics becomes more severe. In turn, we see that the QSL bound is suppressed as time increases [see Fig.~\ref{fig:FIG03}(c)].


\section{UQE, QSL, and many-body systems}
\label{sec:00000000011}

The unified entropy proved to be a useful quantity for witnessing entanglement in multipartite systems~\cite{Sanders_JPhysMath,SciRep_7_1122_2017,Kim_SciRep_8_2018,YYetal_LaserPhysLett_18_2021}. Motivated by the discussion in Secs.~\ref{sec:00000000004} and~\ref{sec:00000000005}, here we investigate the interplay of quantum correlations and speed limits in the nonunitary dynamics of subsystems of many-body systems. We point out that Ref.~\cite{PhysRevA.104.032417} addressed the link between QSL and geometric measure of entanglement for a quantum system evolving unitarily. In addition, Ref.~\cite{PhysRevX.9.011034} provided a geometric lower bound on the QSL for driven ground states of many-body control Hamiltonians. Recent works include the study of QSLs in many-body systems addressing the Kibble-Zurek mechanism~\cite{PhysRevResearch.2.032020}, orthogonality catastrophe~\cite{PhysRevLett.124.110601}, and the dynamics of ultracold gases~\cite{PhysRevLett.126.180603}.

Let ${\mathcal{H}_A}\otimes{\mathcal{H}_B}$ be a finite-dimensional closed quantum system that can be split into two subsystems, ${\mathcal{H}_A}$ and $ {\mathcal{H}_B}$, with dimensions ${d_A} = \dim {\mathcal{H}_A}$ and ${d_B} =  \dim {\mathcal{H}_B}$, respectively. The time-independent Hamiltonian of the whole system $A + B$ is given by $H ={H_A}\otimes{\mathbb{I}_B} + {\mathbb{I}_A}\otimes{H_B} + {H_{AB}}$, where operators $H_A$ and $H_B$ act on their respective subspa\-ces ${\mathcal{H}_A}$ and ${\mathcal{H}_B}$, while ${H_{AB}}$ is the interacting term of the Hamiltonian. The system $A + B$ is ini\-tia\-li\-zed at the state $\rho_0$, which can be chosen either a pure or mixed state, and undergoes the global unitary evolution ${{\mathcal{E}}_t} ({\rho_0}) = {e^{- i t H}} {\rho_0} \, {e^{i t H}}$. The marginal states are given by ${\rho^{A,B}_t} = {\text{Tr}_{B,A}}({{\mathcal{E}}_t} ({\rho_0}))$, with the time-dependent reduced nonunitary dynamics
\begin{equation}
\label{eq:00000000000022}
\frac{d{\rho_t^{A,B}}}{dt}= -i \, {\text{Tr}_{B,A}}([H,{{\mathcal{E}}_t} ({\rho_0})]) ~.
\end{equation}
Without loss of generality, we will focus on the time-dependent reduced state ${\rho^A_t}$, and investigate the dynamic behavior of the family of entropies ${{\text{E}}_{\alpha,\mu}}({\rho^A_t})$ related to subsystem $A$. In this case, applying the results discussed earlier in Sec.~\ref{sec:00000000004}, the rate of change of UQE of ${\rho^A_t}$ satisfies the upper bound
\begin{equation}
\label{eq:00000000000023}
\left| \frac{d}{dt}{{\text{E}}_{\alpha,\mu}}({\rho^A_t}) \right| \leq {h_{\alpha}}({\rho^A_t}){{\left\| \frac{d{\rho^A_t}}{dt} \right\|}_1} ~,
\end{equation}
with the auxiliary function ${h_{\alpha}}(\bullet)$ defined in Eq.~\eqref{eq:0000000000007}. We point out that the bound on the time derivative of UQE in Eq.~\eqref{eq:00000000000023} can be recast as
\begin{equation}
\label{eq:00000000000024}
\left| \frac{d}{dt}{{\text{E}}_{\alpha,\mu}}({\rho^A_t}) \right| \leq 2 \, {h_{\alpha}}({\rho^A_t}) \, \Delta{H} ~,
\end{equation}
where we have used that the Schatten speed of the reduced state $\rho_t^A$ fulfills the following bounds
\begin{equation}
\label{eq:00000000000025}
{\left\| \frac{d{\rho_t^A}}{dt} \right\|_1} \leq {\|[H,{\rho_0}]\|_1} \leq 2 \sqrt{F({\rho_0})}  \leq 2 \, \Delta{H} ~,
\end{equation}
with $F({\rho_0})$ being the quantum Fisher information (QFI), and ${(\Delta{H})^2} = \text{Tr}({\rho_0}{H^2}) - {\text{Tr}({\rho_0}H)^2}$ is the variance of $H$. The proof of Eq.~\eqref{eq:00000000000025} is as follows. For a given operator $\mathcal{O} \subset {\mathcal{H}_A}\otimes{\mathcal{H}_B}$, it has been proved that ${\|{\text{Tr}_B}(\mathcal{O})\|_p} \leq {d_B^{\, (p - 1)/p}}{\|\mathcal{O}\|_p}$~\cite{PhysRevA.78.012308,Rastegin2012}. In particular, by choosing $p = 1$ and $\mathcal{O} = -i [H,{{\mathcal{E}}_t} ({\rho_0})]$, one gets the upper bound ${\left\| {d{\rho_t^A}}/{dt} \right\|_1} \leq {\|[H,{\rho_0}]\|_1} $, where we have applied Eq.~\eqref{eq:00000000000022}, and used the fact that the Schatten $1$-norm is unitarily invariant, i.e., $ {\|[H,{{\mathcal{E}}_t} ({\rho_0})]\|_1} = {\|[H,{\rho_0}]\|_1} $. This result can also be recast in terms of the QFI via the inequality ${\|[H,{\rho_0}]\|_1^2} \leq 4 F({\rho_0})$~\cite{PhysRevX.6.041044,PhysRevA.97.022109}. In addition, QFI satisfies the upper bound $F({\rho_0}) \leq {(\Delta{H})^2}$~\cite{SYu_1302.5311,PhysRevA.87.032324,GToth_1701.07461}, with equality hol\-ding for initial pure states. Finally, by bringing together these results, one readily arrives at the chain of inequalities in Eq.~\eqref{eq:00000000000025}.

Next, to obtain an upper bound on UQE of state $\rho_t^A$, we integrate Eq.~\eqref{eq:00000000000024} over the interval $t \in [0,\tau]$, which yields
\begin{equation}
\label{eq:00000000000026}
\left| {{\text{E}}_{\alpha,\mu}}({\rho^A_{\tau}}) -  {{\text{E}}_{\alpha,\mu}}({\rho^A_0}) \right| \leq 2\, \tau \Delta{H} \,{\left\langle\left\langle {h_{\alpha}}[{\kappa_{\text{min}}}({\rho^A_t})] \right\rangle\right\rangle_{\tau}} ~,
\end{equation}
where we have used that $H$ is time-independent. It is noteworthy that Eq.~\eqref{eq:00000000000026} provides a bound on UQE of initial and final states in terms of the quantum fluctuations of the Hamiltonian. In detail, Eq.~\eqref{eq:00000000000026} means that the entanglement witnessed by UQE is upper bounded by the variance of the time-independent Hamiltonian $H$. In addition, it is also related to the time average of the smallest eigenvalue of the marginal state $\rho_t^A$. Importantly, Eq.~\eqref{eq:00000000000026} implies the lower bound $\tau \geq {\tau^{\text{QSL}}_{\alpha,\mu}}$, where we introduce the QSL time
\begin{equation}
\label{eq:00000000000027}
{\tau^{\text{QSL}}_{\alpha,\mu}} := \frac{\left| {{\text{E}}_{\alpha,\mu}}({\rho^A_{\tau}}) - {{\text{E}}_{\alpha,\mu}}({\rho^A_0})  \right|}{ 2\, \Delta{H} \, {\left\langle\left\langle {h_{\alpha}}[{\kappa_{\text{min}}}({\rho^A_t})] \right\rangle\right\rangle_{\tau}} } ~.
\end{equation}
In order to discuss the tightness of Eqs.~\eqref{eq:00000000000026} and~\eqref{eq:00000000000027}, it is natural to define the relative error
\begin{equation}
\label{eq:00000000000028}
{\delta_{\alpha,\mu}}(\tau) = 1 - \frac{\left| {{\text{E}}_{\alpha,\mu}}({\rho^A_{\tau}}) - {{\text{E}}_{\alpha,\mu}}({\rho^A_0})  \right|}{ 2\, \Delta{H} \, {\int_0^{\tau}} dt \, {h_{\alpha}}[{\kappa_{\text{min}}}({\rho^A_t})]  } ~.
\end{equation}

\begin{figure*}[!t]
\begin{center}
\includegraphics[scale=0.65]{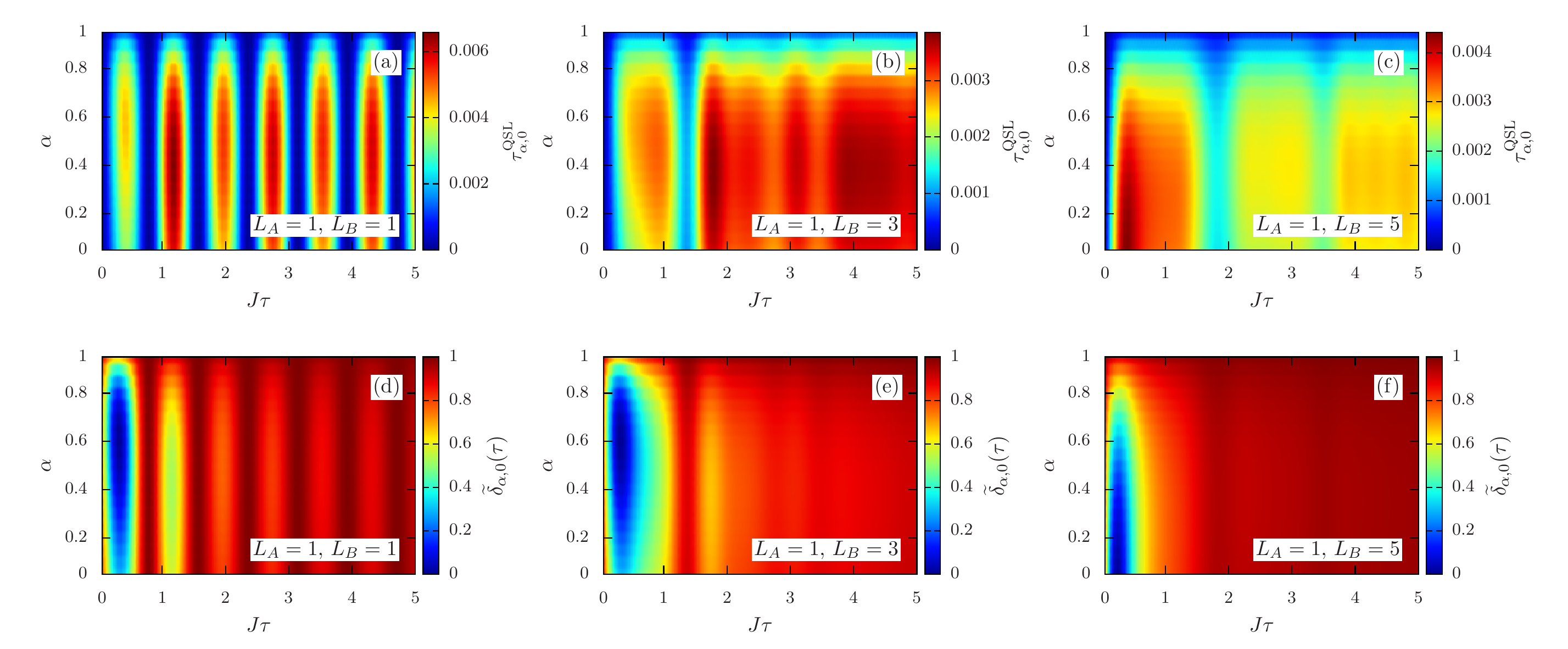}
\caption{(Color online) Plot of the QSL time $\tau^{\,\text{QSL}}_{\alpha,0}$ [see Eq.~\eqref{eq:00000000000027}], and the relative error ${\widetilde{\delta}_{\alpha,0}}(\tau)$ [see Eq.~\eqref{eq:00000000000028}], for the XXZ model, with $J\Delta = 0.5$ [see Eq.~\eqref{eq:00000000000029}]. The system is initialized at the state ${\rho_0} = ((1 - p)/d)\mathbb{I} + p|{\Psi}\rangle\langle{\Psi}|$, with $d = {2^L}$, where $|{\Psi}\rangle = |1,0,1,0,\ldots,1,0\rangle$ is the N\'{e}el state, while $|0\rangle$ and $|1\rangle$ stand for the spin-up and -down state, respectively. Here, we set the system sizes $L = \{2,4,6\}$, with open boun\-da\-ry conditions, while $L_A = 1$, and $L_B = \{1,3,5\}$, also fixing the mixing parameter $p = 0.5$.}
\label{fig:FIG05}
\end{center}
\end{figure*}

We point out that Eq.~\eqref{eq:00000000000027} is the third main result of the paper. Note that the QSL ${\tau^{\text{QSL}}_{\alpha,\mu}}$ is inversely proportional to the variance of the Hamiltonian $H$, and thus fits into the Mandelstamm-Tamm class of QSLs. In addition, by means of the unified entropy, Eq.~\eqref{eq:00000000000027} unveils the role of the correlations into the time it takes to evolve $\rho^A_0$ to an entangled state $\rho^A_t$, for all $t \in [0,\tau]$. We see that ${\tau^{\text{QSL}}_{\alpha,\mu}} > 0$ if a nonzero amount of entanglement is created in finite time during the evolution of the two subregions of the quantum system. However, when no correlation signature is captured by the figure of merit, i.e., for ${{\text{E}}_{\alpha,\mu}}({\rho^A_{\tau}}) \approx {{\text{E}}_{\alpha,\mu}}({\rho^A_0})$, the lower bound on the time of evolution approaches zero ${\tau^{\text{QSL}}_{\alpha,\mu}} \approx 0$. In particular, for $\alpha \rightarrow 1$, we find that Eq.~\eqref{eq:00000000000027} becomes zero, and in this case we find that the von Neumann entanglement entropy is related to a vanishing QSL. It is worth noting that, by means of the quantum fidelity paradigm, Ref.~\cite{PhysRevX.9.011034} addresses the QSL for driven ground states of many-body control Hamiltonians for unitary evolutions. In contrast, by setting the unified quantum entropy as a useful figure of merit, here we discuss the nonunitary dynamics of marginal states of a given quantum many-body system. Importantly, our approach takes in account general input states rather than those states belonging to the ground state manifold of the many-body quantum system.


\subsection{Example}
\label{sec:00000000012}

In the following we illustrate our findings by focusing on the spin-$1/2$ XXZ model with $L$ sites and open boun\-da\-ry conditions
\begin{equation}
\label{eq:00000000000029}
H = {J}{\sum_{j = 1}^{L - 1}}\left( {\sigma_j^x}{\sigma_{j + 1}^x} + {\sigma_j^y}{\sigma_{j + 1}^y} + \Delta\, {\sigma_j^z}{\sigma_{j + 1}^z}\right) ~,
\end{equation}
where $J$ is the exchange coupling constant, and $\Delta$ is the anisotropy parameter. This model has an exact solution via the Bethe ansatz, and its ground state exhibits three phases: a gapless Luttinger liquid phase for $-1 < \Delta \leq 1$; and two gapped phases with long-range order: a N\'{e}el phase for $\Delta > 1$, and a ferromagnetic phase for $\Delta \leq -1$~\cite{takahashi_1999,giamarchi_2003}. Here we consider the initial mixed state ${\rho_0} = ((1 - p)/d)\mathbb{I} + p|{\Psi}\rangle\langle{\Psi}|$, with the N\'{e}el state $|{\Psi}\rangle = |1,0,1,0,\ldots,1,0\rangle$, where $d = {2^L}$ and $0 \leq p \leq 1$, and $\{ |0\rangle, |1\rangle\}$ denote spin-up and -down states, respectively. In this regard, one gets the variance ${(\Delta{H})^2} = {J^2}(L - 1)[2(1 + p) + (1 - p)(1 + (L - 1)p){\Delta^2}]$ for the XXZ model. In the limit of larger system sizes, $L \rightarrow \infty$, the variance will scale with the size of the system.

In Fig.~\ref{fig:FIG05}, we show plots of the QSL time in Eq.~\eqref{eq:00000000000027} [see Figs.~\ref{fig:FIG05}(a)--\ref{fig:FIG05}(c)], and the relative error in Eq.~\eqref{eq:00000000000028} [see Figs.~\ref{fig:FIG05}(d)--\ref{fig:FIG05}(f)], for the XXZ model with open boun\-da\-ry conditions, respective to the R\'{e}nyi entropy ($\mu = 0$). The results for the Tsallis entropy ($\mu = 1$) are qualitatively similar to the case discussed here. Here we choose $J\Delta = 0.5$, the mixing parameter value $p = 0.5$, and set the system sizes $L = \{2,4,6\}$, with $L_A = 1$ and $L_B = \{1,3,5\}$. Note that the system $A+B$ undergoes a unitary dynamics under the time-independent XXZ Hamiltonian, with the subsystem $A$ standing as a two-level system evolving nonunitarily.

In Figs.~\ref{fig:FIG05}(a) and~\ref{fig:FIG05}(d), one verifies that both the QSL time and the re\-la\-tive error are oscillating, time-dependent functions, with the system size $L_A = 1$ and $L_B = 1$. Figure~\ref{fig:FIG05}(a) shows that, for $0 < \alpha < 1$, the QSL exhibits recurrences as a function of time, also presenting regions in which the bound suddenly va\-ni\-shes. On the other hand, the relative error takes small values around $0 \lesssim J\tau \lesssim 1$, while approaches the unity for $J\tau \gg 1$ as time increases. This means the QSL bound becomes loose at later times of the dynamics of the two-level subsystem.

Figures~\ref{fig:FIG05}(b) and~\ref{fig:FIG05}(e) show the QSL time and the re\-la\-tive error for the system size $L_A = 1$ and $L_B = 3$. In Fig.~\ref{fig:FIG05}(b), we see that the QSL time exhibits nonperiodic oscillations, also displaying some recurrences. Note that, for a fixed value $\alpha \in (0,1)$, the QSL exhibits a revival after suddenly approaching small values around $1 \lesssim J\tau \lesssim 1.6$, also experiencing fluctuations in its amplitude. In Fig.~\ref{fig:FIG05}(e), one finds that the relative error is small at earlier times, with $0 \lesssim J\tau \lesssim 1$. In particular, for $J\tau \gtrsim 1$, one finds that ${\widetilde{\delta}_{\alpha,0}}(\tau) \approx 0$ for all $\alpha \in (0,1)$, which means that both the bound on UQE in Eq.~\eqref{eq:00000000000026}, and the QSL in Eq.~\eqref{eq:00000000000027}, become loose.

Next, Figs.~\ref{fig:FIG05}(c) and~\ref{fig:FIG05}(f) show the QSL time and the relative error for the system size $L_A = 1$ and $L_B = 5$. Figure~\ref{fig:FIG05}(c) shows that, for all $0 < \alpha < 1$ and $J\tau \geq 0$, the QSL time in Eq.~\eqref{eq:00000000000027} exhibits a nonperiodic behavior with decreasing amplitudes as a function of time. In addition, for $0.8 \lesssim \alpha \lesssim 1$ and $J\tau \geq 0$, one finds the QSL smoothly vanishes as time varies. This agrees with the fact that ${\tau^{\text{QSL}}_{\alpha,0}} \approx 0$ in Eq.~\eqref{eq:00000000000027} in the limit $\alpha \rightarrow 1$ recovering the von Neumann entanglement entropy. In Fig.~\ref{fig:FIG05}(f), we find that the relative error approaches unity for most of the time, but presents a small peak around $0 \lesssim J\tau \lesssim 0.5$. This means that the bound on UQE is loose, except at earlier times of the dynamics.

Finally, we include general comments on the QSL and relative error for the XXZ model. The amplitude of both quantities decreases as the system size increases. We find that the QSL is inversely proportional to the variance of the Hamiltonian, i.e., ${\tau^{\text{QSL}}_{\alpha,\mu}} \sim \vartheta({\rho^A_{\tau}})/\Delta H$, where the function $ \vartheta({\rho^A_{\tau}}) = |{{\text{E}}_{\alpha,\mu}}({\rho^A_{\tau}}) - {{\text{E}}_{\alpha,\mu}}({\rho^A_0})| {\left\langle\left\langle {h_{\alpha}}[{\kappa_{\text{min}}}({\rho^A_t})] \right\rangle\right\rangle_{\tau}^{-1}}$ depends on the size $L_A$ of subsystem $A$, which can be taken small ($L_A \sim 1$). For larger $L$, the variance will scale with the system size, which means that the QSL time asymptotically decreases as $L$ grows [see Figs.~\ref{fig:FIG05}(a)--\ref{fig:FIG05}(c)]. This behavior should become more evident in the limit of larger $L$. The relative error in Figs.~\ref{fig:FIG05}(d)--\ref{fig:FIG05}(f) takes small values for $J\tau \lesssim 1$, thus signaling that the QSL is loose at later times. We note that, the smaller the relative error, the tighter the QSL in Eq.~\eqref{eq:00000000000027}. It should be noted that a similar conclusion have been reported in the context of equilibration times of many-body systems, but with a focus on the so-called relative purity as a distinguishability measure of quantum states~\cite{PhysRevA.104.052223}.


\section{Conclusions}
\label{sec:00000000013}

In the present work, we have discussed speed limits based on the unified quantum entropy for finite-dimensional quantum systems undergoing arbitrary nonunitary evolutions. Our main contribution lies on the derivation of a family of QSLs related to the UQE for general nonunitary physical processes. In turn, unified entropy denotes a two-parametric information-theoretic quantifier that gives rise to a broad class of entanglement witnesses~\cite{doi:10.1063.1.2165794_Hu_Ye,Sanders_JPhysMath,SciRep_7_1122_2017,Kim_SciRep_8_2018,YYetal_LaserPhysLett_18_2021}. We consider the rate of change of UQE, and derive an upper bound on this quantity. The bound depends on the smallest eigenvalue of the quantum state, also being a function of the Schatten speed [see Eqs.~\eqref{eq:0000000000006} and~\eqref{eq:0000000000008}]. It is noteworthy that the later quantity induces a natural measure of the speed limit for the nonunitary evolution of the quantum state.

We have further addressed the connection between the quantum speed limit time and the unified quantum entropy. In detail, we have derived a lower bound on the time of evolution for nonunitary physical processes [see Eq.~\eqref{eq:00000000000011}]. We find that, apart from the UQE, the QSL depends on the time average of the smallest eigenvalue of the evolved state, and its Schatten speed. Importantly, this result puts forward the discussion involving QSLs for nonunitary evolutions, and provides a class of entropic quantum speed limits.

We have specialized these results to the case of CPTP maps, and for dissipative systems modeled by non-Hermitian Hamiltonians. On the one hand, the QSL bound depends on the set of Kraus operators related to the quantum channel [see Eq.~\eqref{eq:00000000000014}]. On the other hand, for dissipative systems, the QSL bound is recast in terms of the real and imaginary contributions of the non-Hermitian Hamiltonian [see Eq.~\eqref{eq:00000000000019}]. To illustrate the usefulness of the bound, we consider a single-qubit state evolving under the amplitude-damping channel [see Sec.~\ref{sec:00000000008}], and the nonunitary dynamics dictated by a parity-time-reversal symmetric non-Hermitian Hamiltonian [see Sec.~\ref{sec:00000000010}]. The results suggest that the QSL time is tight at earlier times of the dynamics, but becomes loose as time increases and the smallest eigenvalue of the evolved state approaches zero. It is worth mentioning that the non-Hermitian dynamics can also be recast in terms of the so-called metric operator~\cite{Brody_2013,PhysRevA.100.062118}, a Hermitian and positive-definite matrix that endows the Hilbert space with a nontrivial inner product. This could motivate further studies investigating the connection between the quantum speed limit and the metric operator in non-Hermitian systems.

For closed quantum many-body systems, we have derived an upper bound on UQE respective to the nonunitary dynamics of some marginal state of the system [see Sec.~\ref{sec:00000000011}]. We find that the UQE is bounded from above by the quantum fluctuations of the local multiparticle Hamiltonian [see Eq.~\eqref{eq:00000000000026}]. The result applies to many-body systems in which their subsystems are either weakly or strongly coupled. The QSL time is nonzero if the local nonunitary evolution creates a nonzero amount of entanglement in finite time [see Eq.~\eqref{eq:00000000000027}]. In addition, we find the QSL time is inversely proportional to the variance of the Hamiltonian of the system, and thus fits in the Mandelstam-Tamm class of QSLs. We have discussed the QSL time for the integrable XXZ model, thus verifying that the bound remains tight at earlier times, but asymptotically decreases as we increase the system size [see Sec.~\ref{sec:00000000012}].

As a final remark, one can generalize the present discussion in terms of the unified $(\alpha,\mu)$-relative entropy~\cite{IntJTheorPhys_50_1282}, thus obtaining tighter QSLs, and this is an issue that we hope to address in further investigations. Furthermore, given the link between speed limit and geometric measure of entanglement for unitary evolutions~\cite{PhysRevA.104.032417}, one could investigate the trade-off among QSLs, UQE, and geometric measures of entanglement for quantum systems undergoing general physical processes. Finally, the results in this paper could find applications in the subjects of equilibration of many-body systems~\cite{Linden_2010,Short_2011,Eisert_Gogolin_2016,TRO_2018,PhysRevA.104.052223}, noisy quantum metrology~\cite{PhysRevLett.116.120801,Benatti_2014}, and the study of nonequilibrium thermodynamics of dissipative systems~\cite{PhysRevA.92.062128,Ottinger_2011}.


\begin{acknowledgments}
The author would like to acknowledge T. Macr\`{i}, D. O. Soares-Pinto, and F. B. Brito for fruitful discussions. This work was supported by the Brazilian ministries MEC and MCTIC, and the Brazilian funding agencies CNPq, and Coordena\c{c}\~{a}o de Aperfei\c{c}oamento de Pessoal de N\'{i}vel Superior--Brasil (CAPES) (Finance Code 001).
\end{acknowledgments}

\setcounter{equation}{0}
\setcounter{table}{0}
\setcounter{section}{0}
\numberwithin{equation}{section}
\makeatletter
\renewcommand{\thesection}{\Alph{section}} 
\renewcommand{\thesubsection}{\Alph{section}.\arabic{subsection}}
\def\@gobbleappendixname#1\csname thesubsection\endcsname{\Alph{section}.\arabic{subsection}}
\renewcommand{\theequation}{\Alph{section}\arabic{equation}}
\renewcommand{\thefigure}{\arabic{figure}}
\renewcommand{\bibnumfmt}[1]{[#1]}
\renewcommand{\citenumfont}[1]{#1}

\section*{Appendix}


\section{Bounding $\alpha$-purity}
\label{sec:00000000014}

In this appendix, we provide details in the derivation of Eq.~\eqref{eq:0000000000005}. Let ${\rho_t}$ be a full rank, invertible density matrix, thus satisfying the following pro\-per\-ties: (i) ${\rho_t^{\dagger}} = {\rho_t}$, (ii) ${\rho_t} \geq 0$, (iii) $\text{Tr}(\rho_t) = 1$, and (iv) $0 < \text{Tr}({\rho_t^2}) < 1$, for all $t \geq 0$. In this case, by means of contour integration, it has been proved that the following integral representation ${\rho_t^{\alpha}} = {\pi^{-1}}\,{\sin(\pi\alpha)}\,{\int_0^{\infty}} \, {du}\, {u^{\alpha - 1}} \, {({\rho_t} + u\,\mathbb{I})^{-1}} {\rho_t}$ is valid for for all $0 < \alpha < 1$, where $\mathbb{I}$ denotes the identity matrix~\cite{Bathia_Rajendra,ZhangFei_2014_JPA}. Using this result, the $\alpha$-purity ${f_{\alpha}}(\rho_t) = \text{Tr}(\rho_t^{\alpha})$ can recast as follows
\begin{equation}
\label{eq:00000001A}
{f_{\alpha}}({\rho_t}) = \frac{\sin(\pi\alpha)}{\pi}\,{\int_0^{\infty}} \frac{du}{u^{1 - \alpha}} \, \text{Tr}\left({({\rho_t} + u\,\mathbb{I})^{-1}} {\rho_t}\right) ~.
\end{equation}
Starting from Eq.~\eqref{eq:00000001A}, we argue that the absolute value of the time derivative of the $\alpha$-purity satisfies the upper bound
\begin{align}
\label{eq:00000002A}
&\left| \frac{d}{dt}{f_{\alpha}}({\rho_t}) \right| \leq \frac{\sin(\pi\alpha)}{\pi} \left\{ {\int_0^{\infty}} \frac{du}{u^{1 - \alpha}} \, \left|\text{Tr}\left({({\rho_t} + u\,\mathbb{I})^{-1}} \frac{d{\rho_t}}{dt} \right)\right| \right. \nonumber\\
&\left. + {\int_0^{\infty}} \frac{du}{u^{1 - \alpha}} \, \left|\text{Tr}\left({({\rho_t} + u\,\mathbb{I})^{-1}} {\rho_t} {({\rho_t} + u\,\mathbb{I})^{-1}}\frac{d{\rho_t}}{dt}\right)\right|\, \right\} ~,
\end{align}
where we have applied the triangle inequalities $|{A_1} + {A_2}| \leq |{A_1}| + |{A_2}|$ and $|\int du\,  g(u)| \leq \int du |g(u)|$, and used the identity $d{\varrho_t^{-1}}/dt = - {\varrho_t^{-1}} (d{\rho_t}/dt) \,{\varrho_t^{-1}}$, with $\varrho_t = {\rho_t} + u\,\mathbb{I}$. By using the relation ${\rho_t}\, {\rho_t^{-1}} = {\rho_t^{-1}}{\rho_t}$, which holds for any invertible density matrix, for all $t \geq 0$, we obtain
\begin{align}
\label{eq:00000003A}
{\rho_t} {({\rho_t} + u\,\mathbb{I})^{-1}} &=  {({\rho_t^{-1}}({\rho_t} + u\,\mathbb{I}))^{-1}} \nonumber\\
&= {({\rho_t} + u\,\mathbb{I})^{-1}}{\rho_t} ~.
\end{align}
In this regard, substituting Eq.~\eqref{eq:00000003A} into Eq.~\eqref{eq:00000002A}, one readily gets
\begin{align}
\label{eq:00000004A}
&\left| \frac{d}{dt}{f_{\alpha}}({\rho_t}) \right| \leq \frac{\sin(\pi\alpha)}{\pi} \left\{ {\int_0^{\infty}} \frac{du}{u^{1 - \alpha}} \, \left|\text{Tr}\left({({\rho_t} + u\,\mathbb{I})^{-1}} \frac{d{\rho_t}}{dt} \right)\right| \right. \nonumber\\
&\left. + {\int_0^{\infty}} \frac{du}{u^{1 - \alpha}} \,\left|\text{Tr}\left({({\rho_t} + u\,\mathbb{I})^{-1}}{({\rho_t} + u\,\mathbb{I})^{-1}} {\rho_t} \frac{d{\rho_t}}{dt} \right)\right|\, \right\} ~.
\end{align}

In general, given the Schatten $p$-norm ${\| {\hat{\mathcal{O}}}  \|_p} := (\text{Tr}\,[( {\mathcal{O}^{\dagger}}\mathcal{O})^{p/2}])^{1/p}$, the following matrix inequalities hold $|\text{Tr}({\mathcal{O}_1}{\mathcal{O}_2})| \leq {\|{\mathcal{O}_1}\|_{\infty}}{\|{\mathcal{O}_2}\|_1}$, and $|\text{Tr}({\mathcal{O}_1}{\mathcal{O}_2}{\mathcal{O}_3}{\mathcal{O}_4})| \leq {\|{\mathcal{O}_1}\|_{\infty}}{\|{\mathcal{O}_2}\|_{\infty}}{\|{\mathcal{O}_3}\|_1}{{\|\mathcal{O}_4}\|_1}$. Hence, applying these inequalities into the right-hand side of Eq.~\eqref{eq:00000004A} yields
\begin{align}
\label{eq:00000005A}
&\left| \frac{d}{dt}{f_{\alpha}}({\rho_t}) \right| \leq \frac{\sin(\pi\alpha)}{\pi}\left\{ {\int_0^{\infty}} \frac{du}{u^{1 - \alpha}} {\| {({\rho_t} + u\,\mathbb{I})^{-1}}\|_{\infty}} \right. \nonumber\\
&\left. + {\int_0^{\infty}} \frac{du}{u^{1 - \alpha}} {\|{({\rho_t} + u\,\mathbb{I})^{-1}}\|^2_{\infty}} \right\} {\left\|\frac{d{\rho_t}}{dt}\right\|_1} ~,
\end{align}
where we have also used the identity ${\|{\rho_t}\|_1} = 1$, which follows from properties (i)--(iii). Next, by using that ${\|{({\rho_t} + u\,\mathbb{I})^{-1}}\|_{\infty}} = {({\kappa_{\text{min}}}({\rho_t}) + u)^{-1}}$, where ${\kappa_{\text{min}}}({\rho_t})$ sets the minimum eigenvalue of the density matrix $\rho_t$, we see that Eq.~\eqref{eq:00000005A} becomes
\begin{align}
\label{eq:00000006A}
&\left| \frac{d}{dt}{f_{\alpha}}({\rho_t}) \right| \leq \frac{\sin(\pi\alpha)}{\pi} \left\{ {\int_0^{\infty}} du \, \frac{u^{\alpha - 1}}{{\kappa_{\text{min}}}({\rho_t}) + u} \right. \nonumber\\
&\left. + {\int_0^{\infty}} du \, \frac{u^{\alpha - 1}}{({\kappa_{\text{min}}}({\rho_t}) + u)^2}\right\} {\left\|\frac{d{\rho_t}}{dt}\right\|_1} ~.
\end{align}
Finally, by evaluating the integrals in Eq.~\eqref{eq:00000006A}, we obtain the upper bound as follows
\begin{align}
\label{eq:00000007A}
\left| \frac{d}{dt}{f_{\alpha}}({\rho_t}) \right| \leq \left({\kappa_{\text{min}}}({\rho_t}) + 1 - \alpha \right) {({\kappa_{\text{min}}}({\rho_t}))^{\alpha - 2}} \, {\left\|\frac{d{\rho_t}}{dt}\right\|_1} ~.
\end{align}
Importantly, we stress that Eq.~\eqref{eq:00000007A} applies for $\rho_t$ being a nonsingular, invertible density matrix, for all $t \geq 0$. 


\section{Single-qubit non-Hermitian dynamics}
\label{sec:00000000015}

In this appendix we present details on the nonunitary dynamics generated by the two-level non-Hermitian Hamiltonian $\widetilde{H} = \vec{u}\cdot\vec{\sigma}$, where $\vec{u} = \{{\varpi},0,{i\eta}\}$, with ${\varpi},\eta\in\mathbb{R}$, and $\vec{\sigma} = \{{\sigma_x},{\sigma_y},{\sigma_z}\}$ is the vector of Pauli matrices. We set the initial single-qubit state $\rho_0 = (1/2)(\mathbb{I} + \vec{r}\cdot\vec{\sigma})$, where $\vec{r} = \{r\sin\theta\cos\phi, r\sin\theta\sin\phi , r\cos\theta \}$, with $r \in [0,1]$, $\theta \in [0,\pi]$ and $\phi \in [0,2\pi [$, while $\mathbb{I}$ is the $2\times 2$ identity matrix. The evolved state is given by $\rho_t = {U_t}{\rho_0}{U_t^{\dagger}}/\text{Tr}({U_t}{\rho_0}{U_t^{\dagger}})$, with ${U_t} = {e^{-i t \widetilde{H}}}$ being a nonunitary operator. 

We shall begin considering the case $\eta < \varpi$. In this case, the evolution operator is given by ${U_t} = \cos(t \sqrt{{\varpi^2} - {\eta^2}} \,)\mathbb{I} - i\sin( t \sqrt{{\varpi^2} - {\eta^2}} \,)(\hat{u}\cdot\vec{\sigma})$, where $\hat{u} = \vec{u}/\sqrt{{\varpi^2} - {\eta^2}}$ is a unit vector. In this case, the evolved state becomes ${\rho_t} = {(4 + 2{c_t})^{-1}} \left(4{\rho_0} + {c_t}\mathbb{I} + {\vec{q}_t}\cdot\vec{\sigma}\right)$, where we define
\begin{align}
\label{eq:00000002B}
{c_t} &= 2\left[ \hat{u}\cdot{\hat{u}^*} + i({\hat{u}^*}\times\hat{u})\cdot\vec{r} - 1\right]{\sin^2}(t \sqrt{{\varpi^2} - {\eta^2}} \,) \nonumber\\ 
& - i\left((\hat{u} - {\hat{u}^*})\cdot\vec{r}\,\right){\sin}(2t \sqrt{{\varpi^2} - {\eta^2}} \,) ~,
\end{align}
and
\begin{align}
\label{eq:00000003B}
{\vec{q}_t} &= \sin(2t \sqrt{{\varpi^2} - {\eta^2}} \,)\left[(\hat{u} + {\hat{u}^*})\times\vec{r} - i(\hat{u} - {\hat{u}^*})\right] \nonumber\\
&+ 2\, {\sin^2}(t \sqrt{{\varpi^2} - {\eta^2}} \,)\left[ (\hat{u}\cdot\vec{r}\,){\hat{u}^*} + ({\hat{u}^*}\cdot\vec{r}\,)\hat{u} \right.\nonumber\\
&\left. - (1 + \hat{u}\cdot{\hat{u}^*})\vec{r} + i({\hat{u}}\times{\hat{u}^*})\right] ~.
\end{align}

Next, for the case $\eta > \varpi$, the evolution operator is written as ${U_t} = \cosh(t \sqrt{{\eta^2} - {\varpi^2}} \,)\mathbb{I} + \sinh(t \sqrt{{\eta^2} - {\varpi^2}} \,)(\hat{u}\cdot\vec{\sigma})$, with the unit vector $\hat{u} = -i \, \vec{u}/\sqrt{{\eta^2} - {\varpi^2}}$. Hence, one can verify that the evolved state is given by ${\rho_t} = {(4 + 2{h_t})^{-1}} \left(4{\rho_0} + {h_t}\mathbb{I} + {\vec{v}_t}\cdot\vec{\sigma}\right)$, with
\begin{align}
\label{eq:00000005B}
{h_t} &= 2\left[ \hat{u}\cdot{\hat{u}^*} + i({\hat{u}^*}\times\hat{u})\cdot\vec{r} + 1\right]{\sinh^2}(t \sqrt{{\eta^2} - {\varpi^2}} \,) \nonumber\\ 
& + \left((\hat{u} + {\hat{u}^*})\cdot\vec{r}\,\right){\sinh}(2t \sqrt{{\eta^2} - {\varpi^2}} \,) ~,
\end{align}
and
\begin{align}
\label{eq:00000006B}
{\vec{v}_t} &= i \sinh(2t \sqrt{{\eta^2} - {\varpi^2}} \,)\left[(\hat{u} - {\hat{u}^*})\times\vec{r} - i(\hat{u} + {\hat{u}^*})\right] \nonumber\\
&+ 2\, {\sinh^2}(t \sqrt{{\eta^2} - {\varpi^2}} \,)\left[ (\hat{u}\cdot\vec{r}\,){\hat{u}^*} + ({\hat{u}^*}\cdot\vec{r}\,)\hat{u} \right.\nonumber\\
&\left. + (1 - \hat{u}\cdot{\hat{u}^*})\vec{r} + i({\hat{u}}\times{\hat{u}^*})\right] ~.
\end{align}

Finally, at the exceptional point $\eta = \varpi$, the Hamiltonian $\widetilde{H} = {\varpi}({\sigma_x} + i{\sigma_z})$ is gapless, and the evolution ope\-ra\-tor becomes ${U_t} = \mathbb{I} - it\varpi({\sigma_x} + i{\sigma_z})$. In this case, the evolved single-qubit state is given by ${\rho_t} = ({1}/{2})(\mathbb{I} + {\vec{\zeta}_t}\cdot\vec{\sigma})$, with ${\vec{\zeta}_t} = \{{\zeta_x},{\zeta_y},{\zeta_z}\}$, and
\begin{align}
{\zeta_x} &= \frac{r\sin\theta\cos\phi}{1 + 2{\varpi}t(r\cos\theta + {\varpi}t(1 + r\sin\theta\sin\phi))} ~,\\
{\zeta_y} &= \frac{r\sin\theta\sin\phi - 2{\varpi}t(r\cos\theta + {\varpi}t(1 + r\sin\theta\sin\phi))}{1 + 2{\varpi}t(r\cos\theta + {\varpi}t(1 + r\sin\theta\sin\phi))} ~,\\
{\zeta_z} &= \frac{r\cos\theta + 2{\varpi}t(1 + r\sin\theta\sin\phi)}{1 + 2{\varpi}t(r\cos\theta + {\varpi}t(1 + r\sin\theta\sin\phi))} ~.
\end{align}



%

\end{document}